\documentclass[universe,article,accept,moreauthors,pdftex]{Definitions/mdpi}

\def\lsim{\;\raise0.3ex\hbox{$<$\kern-0.75em\raise-1.1ex\hbox{$\sim$}}\;}
\def\gsim{\;\raise0.3ex\hbox{$>$\kern-0.75em\raise-1.1ex\hbox{$\sim$}}\;}
\def\Msun{M_\odot}

\def\kms{\rm ~km~s^{-1}}
\def\ml{~\Msun ~\rm yr^{-1}}
\def\mll{\Msun ~\rm yr^{-1}}

\def \kms {\rm ~km~s^{-1}}

\def\ergs{\rm ~erg~s^{-1}}

\def\ecsb{erg cm$^{-2}$ s$^{-1}$ arcsec$^{-2}$ }
\def\ecsb2{erg cm$^{-2}$ s$^{-1}$ arcsec$^{-2}$}

\usepackage{color}

\def\apj{ApJ}
\def\apjl{ApJ Lett.}
\def\mnras{MNRAS}
\def\nat{Nature}

\def\prd{Phys. Rev. D}
\def\prl{Phys. Rev. Lett.}
\def\pre{Phys. Rev. E}
\def\araa{Ann. Rev. Astron. Astrophys.}                
 \def\aap{Astron. Astrophys.}                  
\def\apjs{Astrophys. J. Suppl. Ser.}                  
\def\apjl{ApJ Lett.}                   

\def\apss{Astroph. Space Sci.}

\def\ssr{Space Sci. Rev.}
\def\jcap{J. Cosmol. Astropart. Phys.}
\def\apspr{Sov. Sci. Rev. Sect. E}
\def\nar{New Astron. Rev.}
\def\aapr{ Astron. Astrophys. Rev.}
\firstpage{1}
\pubvolume{8}
\issuenum{1}
\articlenumber{32}
\pubyear{2022}
\copyrightyear{2022}
\externaleditor{{Academic Editors: Galina L. Klimchitskaya, Vladimir M. Mostepanenko and Nazar R. Ikhsanov} 
}
\datereceived{19 November 2021}
\dateaccepted{3 January 2022}
\datepublished{}
\hreflink{https://doi.org/10.3390/universe8010032} 
\Title{Particle Acceleration in Mildly Relativistic Outflows of Fast Energetic Transient Sources}

\TitleCitation{Particle Acceleration in Mildly Relativistic Outflows of Fast Energetic Transient Sources}

\Author{{Andrei Bykov} 
 *\orcidA{}, Vadim Romansky and {Sergei Osipov}}

\AuthorNames{Andrei Bykov, Vadim Romansky and Sergei Osipov}

\AuthorCitation{Bykov, A.; Romansky, V.; Osipov, S.}

\address[1]{%
Ioffe Institute, 194021 St.~Petersburg, Russia; romanskyvadim@gmail.com (V.R.); osm2004@mail.ru (S.O.)}

\corres{\hangafter=1 \hangindent=1.05em \hspace{-0.82em} Correspondence: byk@astro.ioffe.ru}

\abstract{Recent discovery of fast blue optical transients (FBOTs)---a new class of energetic transient
sources---can shed light on the long-standing problem of supernova---long gamma-ray burst connections.
A distinctive feature of such objects is the presence of modestly relativistic outflows which place them
in between the non-relativistic and relativistic supernovae-related events. {Here we present the results of kinetic particle-in-cell and Monte Carlo simulations of particle acceleration and magnetic field amplification by shocks with the velocities in the interval between 0.1 and 0.7~c. These simulations are needed for the interpretation of the observed broad band radiation of FBOTs.} Their fast, mildly to moderately
relativistic outflows may efficiently accelerate relativistic particles. With particle-in-cell
simulations we demonstrate that synchrotron radiation of accelerated relativistic electrons
in the shock downstream may fit the observed
radio fluxes.
At longer timescales, well beyond those reachable within a particle-in-cell approach, our nonlinear
Monte Carlo model predicts that protons and nuclei can be accelerated to petaelectronvolt (PeV)
energies. Therefore, such fast and energetic transient sources can contribute to galactic populations of
high energy cosmic rays.
}


\begin{document}


\section{Introduction}

The time domain astronomy operating now at all  wavebands from radio to gamma rays has provided unique
information on highly energetic processes in transient astrophysical objects such as supernovae (SNe),
gamma-ray bursts (GRBs), fast radio bursts and others. The~pioneer program of a dedicated search for
supernovae in the optical band started by Fritz Zwicky in 1936, which has later allowed, in~particular, obtaining fundamental results on the accelerated expansion of the Universe via spectroscopy
of SN type Ia, is now ongoing with great perspectives~\cite{2020arXiv200403511K}. The~capabilities of
fast and sensitive wide field imaging suggested for the forthcoming Large Synoptic Survey Telescope
(LSST) would allow detecting many thousands of luminous SNe and tidal disruption events (TDEs) per
year as well as studying other types of transient sources~\cite{2009arXiv0912.0201L}.
Together with the LSST, the~currently operating wide field~\cite{2004ApJ...611.1005G} and survey
~\cite{2021arXiv210413267S} X-ray observatories along with the future high energy missions~\cite{2012ExA....34..551G,2021ExA...tmp...82D,2021arXiv210409533M}, gravitational wave and neutrino
observatories will allow revealing the physical nature of various types of energetic space~transients.

\textls[-15]{The fast blue optical transients (FBOTs)
\cite{2014ApJ...794...23D,2019ApJ...872...18M,2020ApJ...895L..23C,2020ApJ...895...49H} are among the
most interesting recent discoveries. Their appearance is somewhat different from most of the
core-collapse SNe~\cite{2014ApJ...794...23D,2020NatAs...4..893N}. Together with the low-luminosity GRBs they possibly belong to the intermediate class of phenomena filling the gap between
non-relativistic SNe and  ``standard'' long duration gamma-ray bursts, and~which could have volumetric
rates well above that of the GRBs~\cite{2002ARA&A..40..387W,2006Natur.442.1014S}.
The duration of both the energy-momentum release from the central engine and the interaction of the anisotropic ejecta with the outer layers of the progenitor star and its circumstellar matter determine
the transient appearance (see, e.g.,~\cite{2016MNRAS.460.1680I}). }

Three recently studied powerful FBOT sources AT2018cow~\cite{2019ApJ...872...18M}, CSS161010
~\cite{2020ApJ...895L..23C} and ZTF18abvkwla~\cite{2020ApJ...895...49H} were characterized by a low
ejected mass and fast outflows. Indeed, Margutti~et~al.~\cite{2019ApJ...872...18M} found that to explain
the fast rise of the optical and radio emission together with the persistent photosphere appearance in
AT2018cow a wide range of velocities in the range from about 0.02~c to 0.2~c is needed, where $c$ is the speed of light. The~aspherical
ejecta with the range of velocities had the estimated mass $\sim$0.1--1 $\Msun$
\cite{2019ApJ...872...18M}. The~long low frequency $\sl{uGMRT}$ radio observations
~\cite{2021ApJ...912L...9N} allowed estimating the shock radius to be $R = \left(6.1-14.4\right) \times
10^{16}$~cm and the speed of the fast ejecta to be above 0.2~c at 257 days after the shock breakout.
The  mass loss rate of the progenitor star was found to be a few times $10^{-6} \ml$ during the period
of 20--50 years assuming the wind velocity $\sim$$1000 \kms$, while it was possibly 100 times higher in
a period of a few years just before the event~\cite{2021ApJ...912L...9N}.

Another interesting FBOT source is CSS161010 located in a dwarf galaxy at a distance about 150~Mpc
~\cite{2020ApJ...895L..23C}. On~the basis of the synchrotron interpretation of its radio emission,
which is peaked at about 100 days after the FBOT event, the~authors suggested a presence of a mildly
relativistic outflow of four-velocity 0.55~c driving a blast wave. They estimated the ejected mass to
be in the range of 0.01--0.1 $\Msun$. The~outflow is faster than that estimated for AT2018cow
and is similar to that in ZTF18abvkwla. The~origin of the bright X-ray luminosity is attributed to an
emission component, which is likely different from the primary one, which produced the synchrotron
radio emission.
Recent radio, millimeter wave and X-ray observations~\cite{2021arXiv211005490H,2021arXiv211005514B}
of a short-duration luminous FBOT transient ZTF20acigmel (AT2020xnd) located at z = 0.2433 indicated a
presence of a fast ejecta with a $\sim$0.2~c speed shock with estimated energy above 10$^{49}$ ergs.
AT2020xnd has shown high radio luminosity of $L_{\nu}\sim 10^{30}$ergs$^{-1}$Hz$^{-1}$ at 20~GHz
almost 75 days after the event~\cite{2021arXiv211005514B}. The~observational data suggested a shock
driven by a fast outflow of velocity 0.1--0.2~c interacting with the dense circumstellar matter shaped by
an intense wind of $\dot M\approx10^{-3}\mll$ with velocity of $v_w=1000$ km s$^{-1}$ from the progenitor
star~\cite{2021arXiv211005514B}; the presence of a steep density profile of $\rho(r) \propto r^{-3}$
in the wind was suggested by~\cite{2021arXiv211005490H}. Similar to AT2018cow, the~detected X-ray
emission is in excess compared to the extrapolated synchrotron spectrum and constitutes a different
emission component, possibly powered by accretion onto a newly formed black hole or neutron~star.

The distinctive features of the four FBOT transients discovered so far are their high peak bolometric
luminosity $L\gsim 10^{43} \ergs$ and a rapid timescale of a few days duration. Multi-wavelength
observations of these objects uncovered the presence of powerful sub- or mildly relativistic outflows
which are likely originated from either rare SN-type or TDE-type events with intermediate or stellar mass
black holes (see, e.g., \cite{2021ApJ...911..104K}). The~kilonova type sources with the neutron stars mergers typically eject the masses in the range of \mbox{10$^{-4}$--10$^{-2} \Msun$} (and even $\sim 0.1~\Msun$ for the black hole---neutron star mergers)
with velocities $\sim$0.1–0.3 c~\cite{2019LRR....23....1M}. It is interesting that the recent model of a
supernova from a primordial population III star of a 55,500 $\Msun$ mass with general relativistic
instability~\cite{2021MNRAS.503.1206M} predicts the ejecta velocities of about 0.3~c.
Earlier mildly and moderately relativistic ejecta outflows were found in a few broad line type Ic SNe
(e.g., \cite{2010Natur.463..513S,2016ApJ...830...42C}).
The geometry and structure of the outflows producing FBOTs depend on the source of their power
and it is a subject of modeling~\cite{2019ApJ...872...18M}. Relativistic mass ejection in spherically
symmetric shock outflows of core-collapse supernovae was studied in detail in~\cite{1989ASPRv...8....1I,1999ApJ...510..379M,2001ApJ...551..946T}. Their high-velocity solutions
demonstrated rather a steep dependence of the deposited kinetic energy $E_k$ as a function of the ejecta
four-speed $E_k \propto (\beta \Gamma)^{-5.2}$. Much flatter energy---ejecta velocity distribution $E_k
\propto (\beta \Gamma)^{-2.4}$ can be obtained for engine-driven asymmetric supernovae with a powerful
activity of compact stellar remnants
~\cite{2001ApJ...550..410M,2012ApJ...750...68L,2014ApJ...797..107M,2021NewAR..9201614C}. {Recently, numerical models of supernova explosions where the supernova ejecta interacts with the relativistic wind from the central engine were constructed in~\cite{Suzuki2017}. The~formation of relativistic flows in the interaction of the powerful non-thermal radiation produced by the central machine with the supernova ejecta was considered in the papers~\cite{Suzuki2019,Suzuki2021}}.

Very luminous optical FBOT events with light curves extending to a couple of weeks can be expected in the
case of shock breakout into a dense circumstellar shell produced by the dense progenitor wind a few years
before the SN event~\cite{1971Ap&SS..10...28G,2011ApJ...729L...6C,2013MNRAS.429.3181T}. The~presence of
an hour timescale central engine activity with a luminosity of about 10$^{47} \ergs$ producing mildly
relativistic jet was proposed in~\cite{2016MNRAS.460.1680I} to model SN 2006aj associated with a
low-luminosity GRB. The~bright high frequency radio emission of FBOTs and, possibly, the~hard X-ray
component detected in some events of this type, is produced by relativistic electrons accelerated by
shocks driven by fast moderately relativistic outflows from the central engine.
Moreover, the~synchrotron self-absorption effects~\cite{1998ApJ...499..810C} are apparent
in some FBOT~spectra.

The SN shock breakout is usually accompanied by a bright ultraviolet (UV) flash, and~it is likely that
afterwards the shock enters a collisionless regime~\cite{2017hsn..book..967W} with the X-ray dominated
spectrum. Some models of relativistic shock breakout which consider a multifluid structure of
a relativistic shock mediated by radiation in a cold electron-proton plasma are currently
under discussion~\cite{2010ApJ...725...63B,2020PhRvE.102f3210L}.
A MeV gamma-ray flash of the total energy $\sim$$10^{48}$ erg lasting from a few seconds to a few hours
was predicted in~\cite{2018MNRAS.476.5453G,2020MNRAS.499.4961I} for some SN events.
At the later SN stages (after a few days from the event) the radiative shock is transforming into a
collisionless plasma shock regulated by kinetic plasma instabilities. A~specific feature of the
collisionless shocks is their ability to create a powerful non-thermal particle population and to
accelerate relativistic particles~\cite{2016RPPh...79d6901M}. The~collisionless shock structure and the
efficiency of particle acceleration depend on the shock speed, on~magnetic field inclination to the shock
normal and on the plasma magnetization parameter~\cite{SironiObliquity,2017SSRv..207..319P}. Modeling of
particle acceleration by non-relativistic shocks of velocities below 0.1~c in supernova remnants was
discussed in~\cite{2013MNRAS.431..415B} while studies of relativistic shocks was presented in~\cite{2018MNRAS.473.2364B}.

Kinetic simulations of the
efficiencies of the shock ram pressure conversion to magnetic fluctuations and relativistic particles
are needed to provide an adequate interpretation of the observed non-thermal radiation.
The mildly relativistic magneto-hydrodynamic (MHD) flows were shown (see, e.g.,
\cite{2009JCAP...11..009L,2012SSRv..173..309B}) to be the most efficient environment providing
the maximum energies of the accelerated nuclei for a given magnetic/kinetic luminosity of the
power engine. Recent models suggested a possibility of cosmic ray acceleration to ultra high energies in
the low-luminosity GRBs associated with SNe (see, e.g.,~\cite{2019PhRvD.100j3004Z,2020ApJ...902..148S})
and in relativistic SNe~\cite{2018SSRv..214...41B}. Fast outflows from SNe with dense circumstellar
shells could accelerate cosmic rays up to the high energy regime on a few weeks timescale (e.g.,
\cite{2016APh....78...28Z,2019ApJ...874...80M}).

The structure and particle acceleration in the fast collisionless shocks can be successfully modeled
with the kinetic particle-in-cell (PIC) technique~\cite{2015SSRv..191..519S,2016RPPh...79d6901M} at many
thousands of the particle gyro-scales. On~the other hand, as~the observed multi-wavelength spectra of
supernova remnants and GRBs compellingly demonstrated that the spectra of accelerated particles extend
to many decades in the particle momentum, a~combination of both microscopic (see Section 
 \ref{PiC}) and
macroscopic kinetic models (see Section \ref{MC}) is necessary to construct realistic models of such sources.
Cosmic ray acceleration by supernova remnants is a subject of extensive modeling~\cite{2013MNRAS.431..415B,2013ApJ...763...47P,2013A&ARv..21...70B,2014IJMPD..2330013A,
2018SSRv..214...41B,2021Univ....7..324C}. One of the most uncertain points is at what stage PeV
regime cosmic rays can be accelerated. We discuss in Section \ref{MC} high energy cosmic ray
acceleration in FBOT-type sources as potential~pevatrons.

{Here we will present 2D PIC simulation of sub and mildly relativistic shocks within the shock speed interval 0.1--0.7~c 
 in proton-electron plasmas which could be applied to FBOT sources modeling. The~fiducial case in our work is $\displaystyle$ 0.3~c.

Earlier, Park~et~al.~\cite{Park15} presented a set of 1D PIC models of shocks in a range of velocities up to 0.1~c. They performed in the 1D case long simulation runs up to $5 \times 10^5 \omega_{pe}^{-1}$. For~higher shock speed of  0.75~c  Crumley~et~al.~\cite{Crumley,Romansky2018} published  2D PIC simulation results. In~the quasi-parallel shock of a velocity 0.75~c simulation of about 4100 $\omega_{pi}^{-1}$ duration allowed them to model Bell-mediated shock. In~particular, they noted that acceleration of electrons is likely initially associated with the shock drift acceleration and then switches to diffusive shock acceleration regime as it was seen in~\cite{Park15}.

To model radio emission observed from CSS161010, where the shock velocity of about  0.3~c was suggested by observations, we make 2D PIC simulations with $m_p/m_e$ ratio of 100, which are limited to the timescale   $\sim$7 $\times~10^4~\omega_{pe}^{-1}$. We discuss possible extrapolations of electron spectra to the energy range needed to model the radio emission.}



\section{Particle-in-Cell Simulations of Fast Mildly Relativistic~Shocks}\label{PiC}

In this section, we present the results of particle-in-cell simulations of mildly relativistic shocks at
gyro-scales with application to the observed non-thermal emission from~FBOTs.

To simulate the structure and non-thermal particle acceleration by a mildly relativistic collisionless
shock wave we have employed the publicly available code Smilei
 developed by Derouillat~et~al.
~\cite{Smilei18}. This code is based on explicit Finite Differences Time Domain approach for solving
Maxwell equations and on a relativistic solver for particle movement with charge-conserving algorithm
proposed by~\cite{Esirkepov}.

For model shock initialization we employ a common approach: a homogeneous plasma flow collides with
an ideal reflecting wall. The~simulation box is two-dimensional, with~the number of cells
$\displaystyle Nx$ = 204,800 in the direction along the flow velocity and $\displaystyle Ny = 400$
in the transverse direction, while particle velocities and the electromagnetic field are represented by
full 3D vectors. Homogeneous plasma flows in through the right boundary and the reflecting wall
is placed at the left boundary. Boundary conditions in the transverse direction are periodic.
The initial magnetic field $\displaystyle\vec{B}$ lies in the plane of the simulation inclined by
the angle $\displaystyle\theta$ to the velocity of the flow. The~electric field is initialized to
compensate the Lorentz force in the laboratory frame $\displaystyle\vec{E} = -\vec{v} \times
\vec{B}/c$. The~velocity of plasma flow $\displaystyle v$  is 0.3~c 
  and its Lorentz-factor $\displaystyle\gamma$ is $\displaystyle 1.05$. The~magnetization $\displaystyle\sigma =
B^2/4 \pi n \gamma m_p c^2$, where $\displaystyle n$ is upstream concentration and $\displaystyle m_p$ is mass of proton, in~all the modeled configurations is about $\displaystyle 10^{-4}$.
The electron mass $\displaystyle m_e$ is artificially increased up to $\displaystyle m_p/m_e = 100$ in order to save
computational resources. All the quantities obtained from the simulation can be scaled with respect to
the plasma concentration, which can be chosen arbitrary during the data analysis. The~spatial grid step
is $\displaystyle dx = 0.2 c/ \omega_{e}$ and the time step is $\displaystyle dt = 0.09~ \omega_{e}^{-1}$,
where $\displaystyle\omega_{e}$ is electron plasma frequency $\displaystyle\omega_e = \sqrt{4 \pi n
e^2/ m_e \gamma}$, $\displaystyle e$ is the
absolute value of the electron charge. Within~such a setting the simulation box size along the $\displaystyle x$-axis corresponds to $\displaystyle 500$
gyroradii of protons in plasma flow $\displaystyle r_g = m_p v c \gamma/e B$ and along the $\displaystyle y$-axis---to 1 gyroradius. The~maximum simulation time is $\displaystyle 7\times 10^4~\omega_{e}^{-1}$
 or
about $\displaystyle 100$ inverse proton gyrofrequencies. For~plasma concentration $\displaystyle n\approx
1$~cm$^{-3}$, this corresponds to timescale about 1 s, which is much smaller than the typical
activity period of~FBOTs.

\begin{figure}[H]
	\includegraphics[width=10.5 cm]{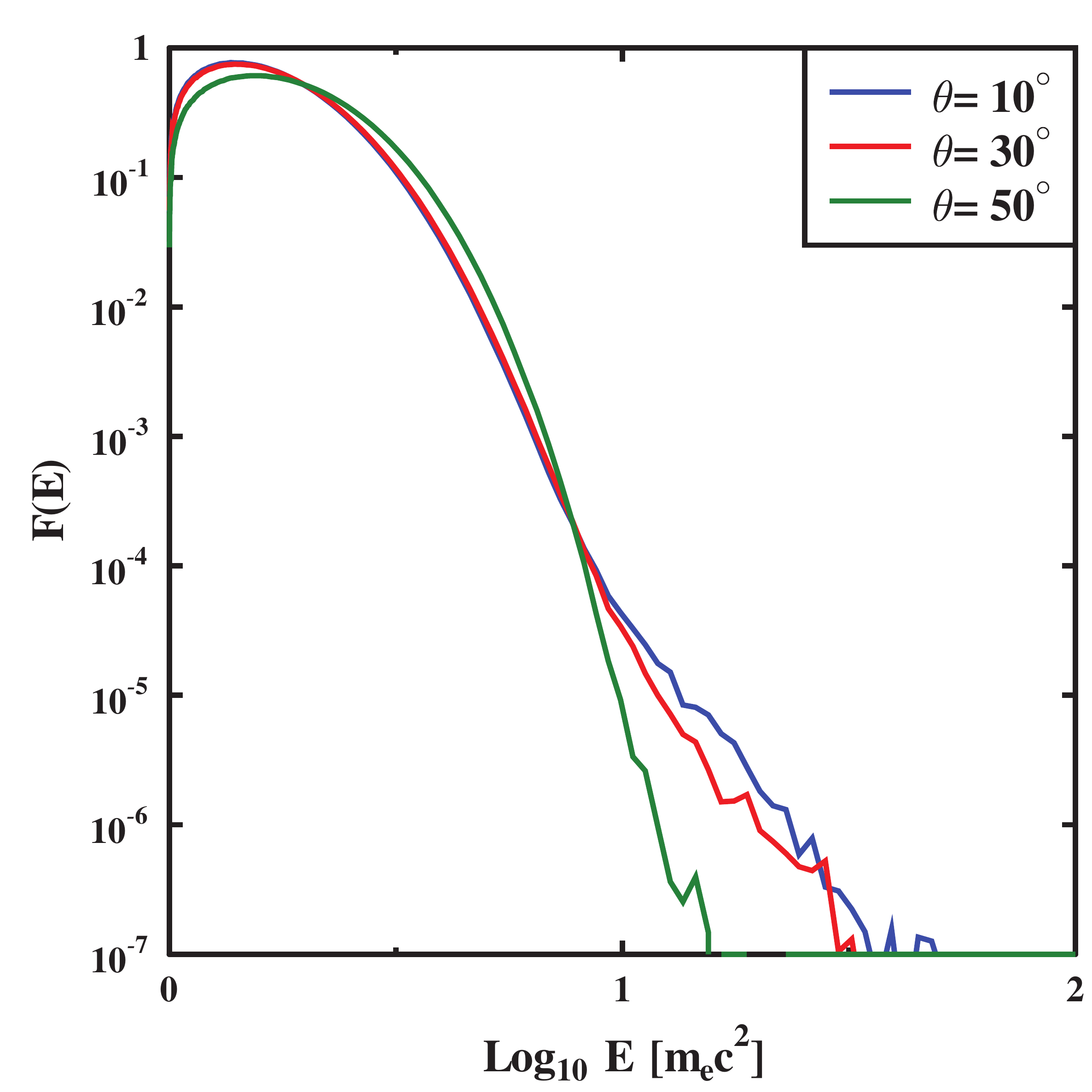}
	\caption{Electron 
 distribution function in the shock downstream with initial parameters \mbox{$v = 0.3~c,$} $
	\sigma = 0.0002$ and different inclination angles.}
	\label{compareelectrons}
\end{figure}

The efficiency of particle acceleration depends on the inclination angle $\displaystyle\theta$,
especially in the case of relativistic flows~\cite{SironiObliquity, 2014ApJ...794..153G, Crumley, Romansky2018}. To~participate in the diffusive acceleration process, a~particle needs to escape from the shock front and
if it moves along the magnetic field, the~maximum velocity along the $\displaystyle x$-axis is $\displaystyle c
\cos(\theta')$, and~it should be larger than the shock velocity $\displaystyle v_{sh}'$ (all quantities
here are measured in the upstream frame). As~one can see in Figure~\ref{compareelectrons}, the~high
energy tail of the electron distribution is much higher for a quasiparallel shock. Such an angle
dependence may lead to the presence of different electron distributions within one object, and~this may
possibly explain the difference between the spectral indices of synchrotron radio and inverse Compton
X-ray radiation observed in FBOT AT2018cow~\cite{2019ApJ...872...18M}.

\begin{figure}[H]
    	    \includegraphics[width=10.5 cm]{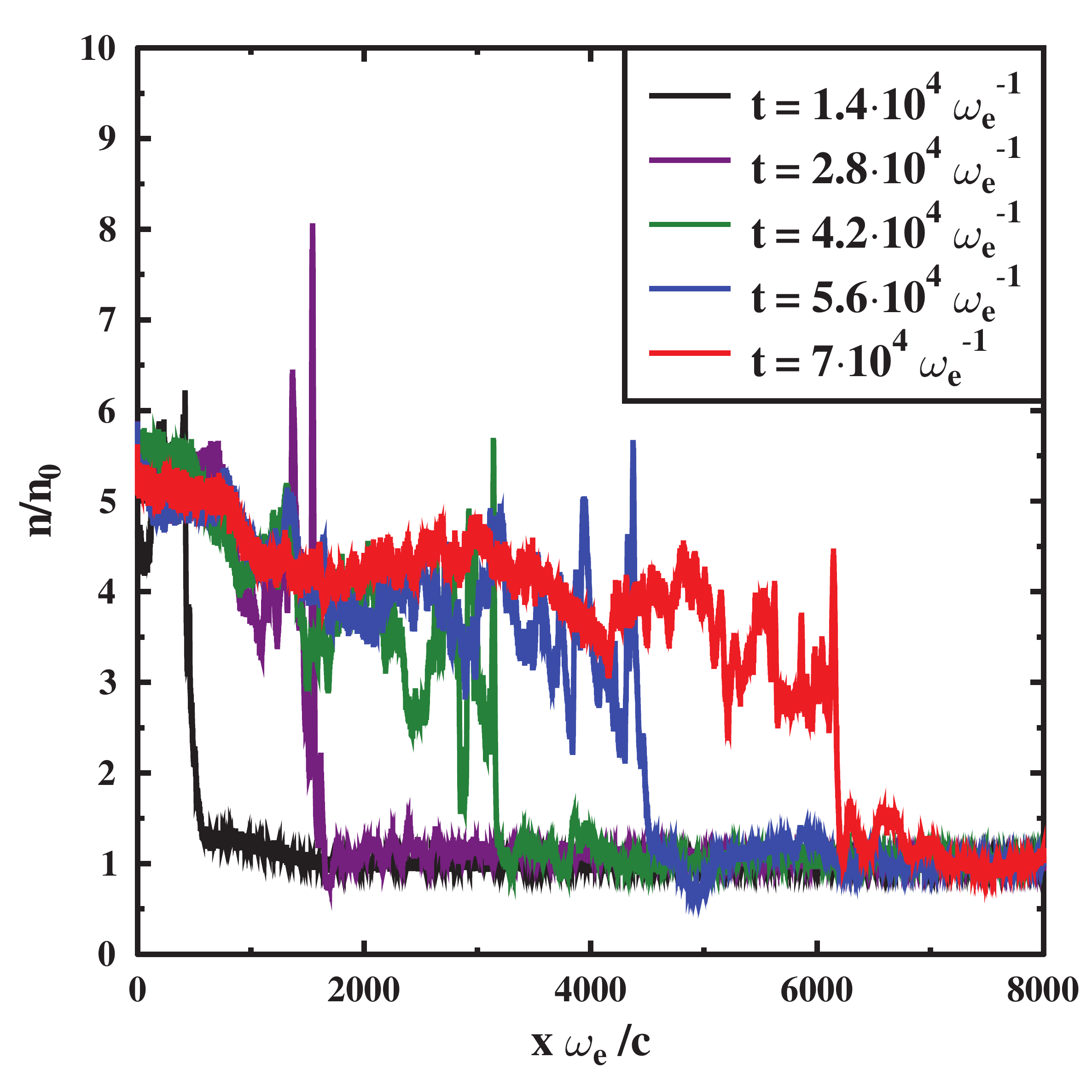}
	    \caption{Time evolution of concentration normalized to the far upstream~concentration.}
	    \label{concentration}
\end{figure}

\begin{figure}[H]
	    \includegraphics[width=0.65\textwidth]{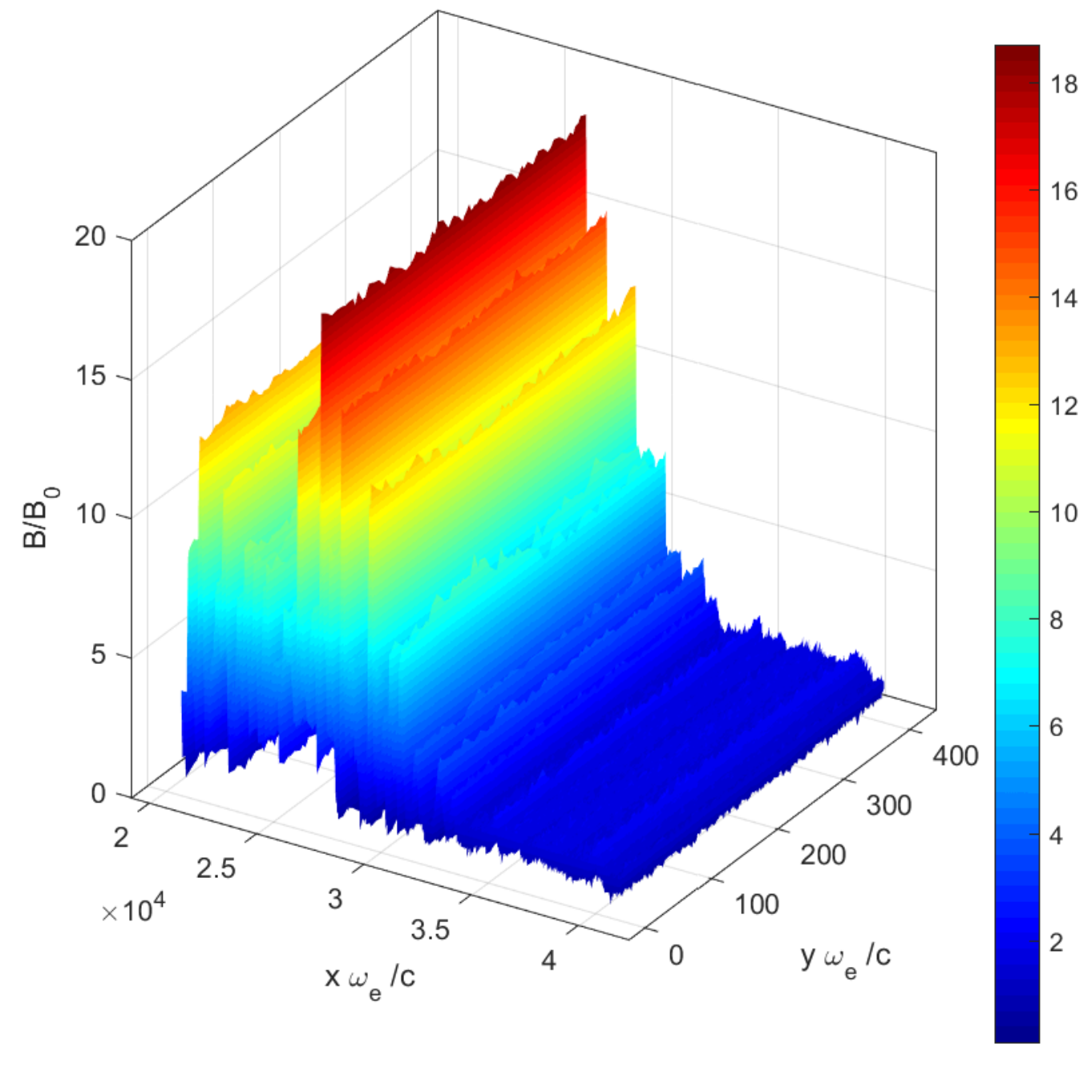}
	    \caption{Magnetic field normalized to the far upstream magnetic~field.}
	    \label{field}
\end{figure}

For further modeling of synchrotron radiation from FBOT CSS161010 we have used a setup with parameters
$\displaystyle v =$ 0.3~c, $\displaystyle \sigma = 0.0002$ and $\displaystyle\theta = 30^{\circ}$ (a quasiparallel
shock). Time evolution of the concentration profile in such a shock averaged in the transverse
direction is shown in Figure~\ref{concentration} and the corresponding magnetic field at the moment
$\displaystyle 7\times 10^{4}~\omega_{e}^{-1}$ is shown in Figure~\ref{field}.

With the computed concentration profile we can determine the coordinate of the shock front:
we consider that $\displaystyle x_{sh}$ is the first point from the right where the concentration is two
times larger than that in the far upstream. The~average shock velocity is the ratio of the shock
coordinate and the simulation time $\displaystyle v_{sh} = x_{sh}/t$. At~later times the shock
wave propagation is close to a stationary regime. The~shock velocity measured in the downstream frame
is close to the constant value $\displaystyle v_{sh} \approx $ 0.085~c. In~the upstream (observer) frame
it corresponds to $\displaystyle v_{sh}'=\left(v_{sh} + v\right)/\left(1 + v_{sh}v/c^2\right) \approx$ 0.38~c.

\vspace{-15pt}

\begin{figure}[H]
	    \includegraphics[width=10.5 cm]{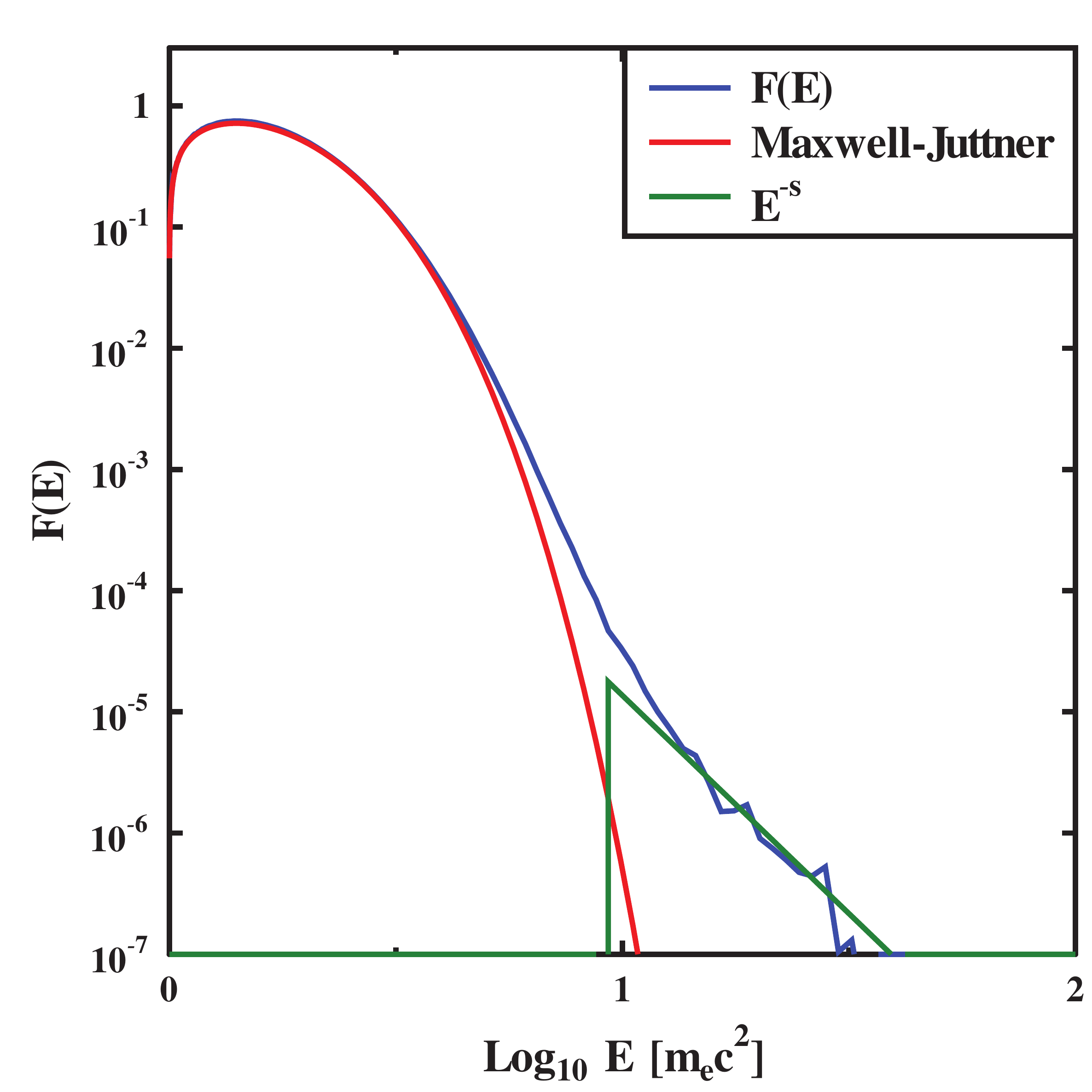}
	    \caption{The 
 electron distribution function in the shock downstream with initial parameters $v = 0.3~c,$ $\sigma = 2\times10^{-4}$ and $\theta = 30^{\circ}$.}
	    \label{electrons}
\end{figure}

\textls[-15]{The simulations have allowed us to compute model distributions of electrons.
In the downstream region such distributions have a complex shape, where one can see
two main components: a thermal peak and a power law tail at high energies.
At low energies ($\displaystyle E < 5 m_e c^2$) we approximate the distribution with a
Maxwell-Juttner function: an iterative process is employed to minimize the
functional $\displaystyle f(T) = \int_{m_e c^{2}}^{5 m_e c^{2}} (F(E) - F_{mj}(E,T))^{2}dE$, where $\displaystyle F(E)$ is simulated electron distribution function and $\displaystyle F_{mj}(E,T)$ is Maxwell-Juttner distribution function, and~find
the effective temperature. For~high energies ($\displaystyle 20~m_e c^{2} < E < 50~m_e c^{2}$) we follow
a least squares approach for linear regression in double logarithmic coordinates and obtain the power law
spectral index. The~electron spectrum in a close downstream of the shock (5000~grid cells behind the
shock) at time $\displaystyle t$ = 70,000~$\omega_{e}^{-1}$ for the setup with $\displaystyle\sigma =
0.0002$ and $\theta = 30^\circ$ and its approximation with temperature $\displaystyle T_e = 5\times
10^{10}$~K and spectral index $\displaystyle s = 3.59$ are shown in Figure~\ref{electrons}.}

\begin{figure}[H]
	    \includegraphics[width=10.5 cm]{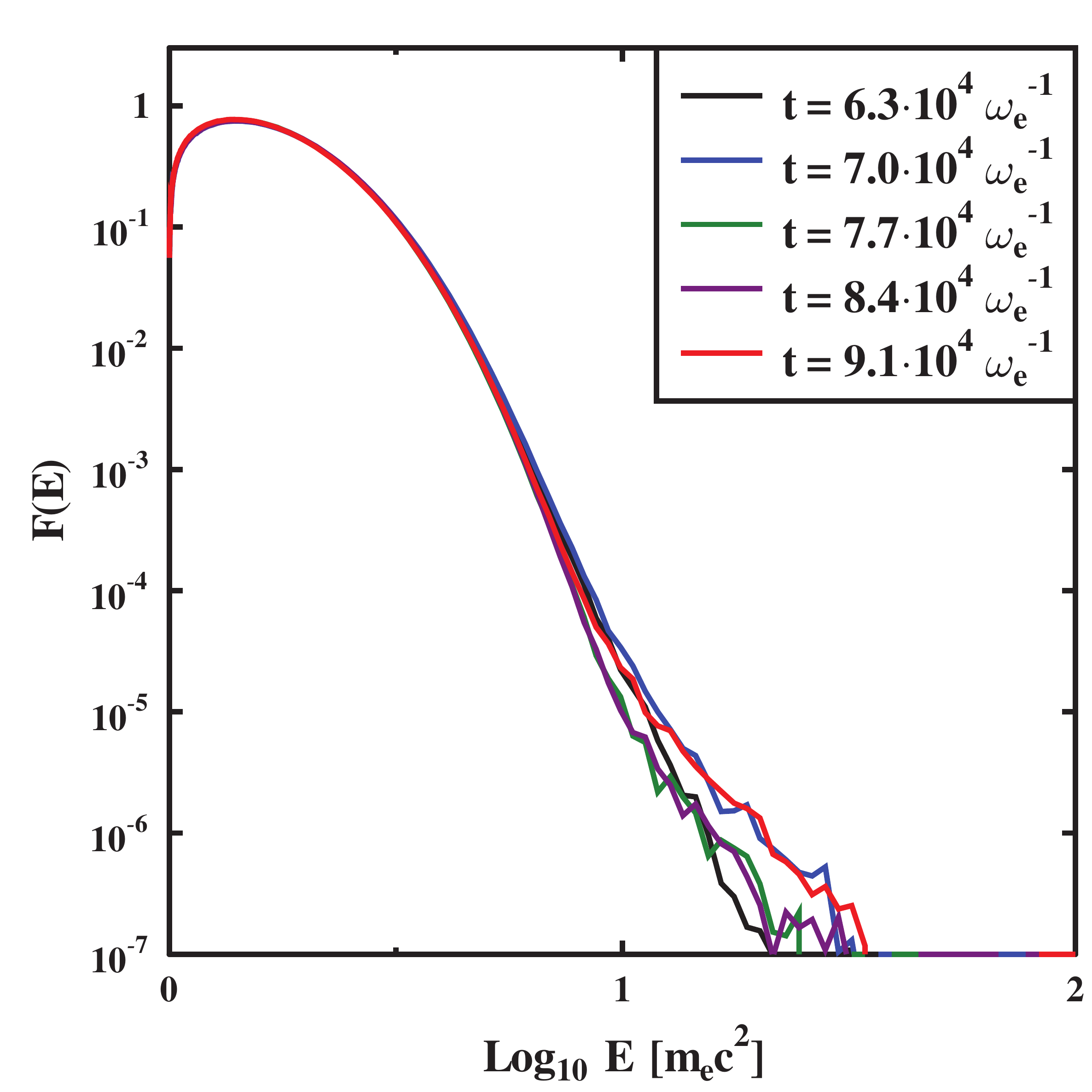}
	    \caption{The 
 electron distribution function in the shock downstream at different time~moments.}
	    \label{electrons_time}
\end{figure}

{Time dependence of electron distribution function at later stages of simulation is shown in Figure~\ref{electrons_time}. Distribution is not stationary and one can see the oscillation of the electron spectral distribution due to the influence of magneto-hydrodynamic instabilities, as~described in~\cite{Crumley, Romansky2021}. So we have chosen the electron distribution at time $\displaystyle t$ = 70,000~$\omega_{e}^{-1}$ for the further modeling.}

Recent observations of FBOTs
~\cite{2014ApJ...794...23D,2019ApJ...872...18M,2020ApJ...895L..23C,2020ApJ...895...49H} discussed above
have shown that their synchrotron spectra are strongly influenced by synchrotron self-absorption. At~the
given time moment, observable radio fluxes rise as $\displaystyle F_{\nu} \propto {\nu}^{5/2}$ for low
frequencies, then have a peak and fade with a power law tail, which depends on the particular electron
distribution function. Additionally, the~time dependence of such spectrum is very specific and its details
could be used to imply a number of source parameters: the magnetic field and the shock radius can be determined from the measured maximum of the light-curve $\displaystyle F_{max}$ at given frequency
$\displaystyle\nu$, once the fractions of energy in accelerated electrons and magnetic field
$\displaystyle\epsilon_{e}$ and $\displaystyle\epsilon_{B}$ are established, with~the relations
derived by Chevalier \citep{1998ApJ...499..810C}.

	\vspace{-15pt}

\vspace{-10pt}

\begin{equation}\label{ChevR}
R = {\left( \frac {6 \epsilon_B {c_6}^{s+5}{F_{max}}^{s+6}{
D}^{2s+12}}{\epsilon_e f \left( s-2 \right) {\pi}^{s+5}{{c_5}}^{s+6}
{E_1}^{s-2}} \right)} ^{ \frac{1}{2s+13} } \frac{2 c_1}{\nu},
\end{equation}
\begin{equation}\label{ChevB}
B = { \left(\frac {{\epsilon_B}^2 36 {\pi}^{3}{c_5}}{{
\epsilon_e}^{2}{f}^{2} \left( s-2 \right) ^{2}{{c_6}}^{3}{{E_1}}^{2 s-4}F_{max}{D}^{2}} \right)}
^{\frac{2}{2s+13}}\frac{\nu}{2 c_1},
\end{equation}
where $\displaystyle s$ is the electron spectral index, derived from the power law tail of emissivity
spectrum and its time dependence, $\displaystyle$ $c_1$ 
, $c_5$, $c_6$ are the Pacholczyk constants~\cite{Pacholczyk}, depending on $\displaystyle s$, $\displaystyle E_{1}$ is the minimum energy of
the electron power law distribution, which usually equals to the electron rest energy, $\displaystyle f$
is the filling factor---the emitting fraction of the source volume, and~$\displaystyle D$ is the distance to the source.
A problem of this method is that it needs a lot of assumptions about the shock structure and electron
distribution. In~this model the emitting region is considered to be a homogeneous flat disk with the
radius $\displaystyle r$ and depth $\displaystyle f \times r$. 
 There is also a model modification for a
spherical source with an account for source inhomogenuity~\cite{2021arXiv211005490H}. The~electron
distribution is considered to be a power law. Additionally, fractions of energy $\epsilon_e$ and
$\displaystyle\epsilon_{B}$ are unknown and are often chosen according to the equipartition rule
$\displaystyle\epsilon_e = \epsilon_B = 1/3$. However the simulation results show, that these
values are unlikely to be that large. Following Chevalier~\cite{2017hsn..book..875C} we define the energy
fractions in terms of the upstream kinetic energy density $\displaystyle (m_p +m_e) n_u {v_{sh}'}^2$,
where $\displaystyle n_{u}$ is the upstream concentration. In~this notation $\displaystyle\epsilon_{e} =
E_e/\rho_u {v_{sh}'}^{2}$ and $\displaystyle\epsilon_{B} = B^{2}/\left(8 \pi (m_p +m_e) n_u {v_{sh}'}^{2}\right)$, $\displaystyle E_{e}$ is the accelerated electron energy evaluated by subtracting from
the total electron energy, that corresponds to the Maxwell--Juttner fit of the distribution function.
The values obtained by particle-in-cell simulations in our fiducial setup are $\displaystyle\epsilon_{e}
= 0.014$ and $\displaystyle\epsilon_B = 0.03$. They depend on the initial conditions, but~for a wide
variety of parameters we see that the fraction of energy in magnetic field is lower than $\displaystyle
10\%$. Particle-in-cell simulations have rather small scales and cannot describe the influence of long
wave upstream instabilities caused by high energy particles, but~Monte Carlo simulations show similar
values, as~described~below.

Using the electron distribution function obtained from the PIC simulation we can evaluate the
spectral density of the energy flux of synchrotron radiation from the source, taking into account
the effect of synchrotron self absorption. Standard formulae for that effect are described in detail in
~\cite{Ghisellini}.
Emitted power per unit frequency per unit volume is
\begin{equation} \label{emission}
I(\nu)=\int_{E_{min}}^{E_{max}} dE \frac {\sqrt {3}{e}^{3}n F(E) B \sin ( \phi)}{{m_e}{c}^{2}}
\frac{\nu}{\nu_c}\int_{\frac {\nu}{\nu_c}}^{\infty }\it K_{5/3}(x)dx,
\end{equation}
where $\phi$ is the angle between the magnetic field and the line of sight, $\displaystyle\nu_{c}$ is the critical frequency
$\displaystyle\nu_{c} = 3 e^{2} B \sin(\phi) E^{2}/4\pi {m_{e}}^{3} c^{5}$, and~$K_{5/3}$ is
the modified Bessel function. The~absorption coefficient is
\begin{equation}\label{absorption}
k(\nu)=\int_{E_{min}}^{E_{max}}dE\frac {\sqrt {3}{e}^{3}}{8\pi m_e \nu^2}\frac{n B\sin(\phi)}{E^2}
\frac{d}{dE} E^2 F(E)\frac {\nu}{ \nu_c}\int_{\frac {\nu}{ \nu_c}}^{\infty }K_{5/3}(x) dx.
\end{equation}

The obtained spectral density is further integrated over the volume of the source, which is described
as a spherical shell, whose volume is determined by a filling factor $\displaystyle f = 0.5$. Hence the
total emissivity and observable flux at distance $\displaystyle D$ can be derived. The~magnetic field
$\displaystyle B$ is considered constant and perpendicular to the line of sight. {The concentration in the stellar wind depends on radius as $\displaystyle n \propto r^{-2}$, but~in the downstream of the shock we assume it constant}. The~electron distribution
function is also constant in the volume of the source.
This modeling is further applied to explain the spectrum of the FBOT transient CSS161010. As~described in
~\cite{2020ApJ...895L..23C} at time $\displaystyle t = 357$ days after explosion its shock velocity was
$v_{sh} = 0.36c$, which corresponds to the modeled upstream plasma flow velocity $\displaystyle vs. =
0.3c$, measured in the downstream frame. At~this moment, parameters derived with Equations~(\ref{ChevR})
and (\ref{ChevB}), assuming the equipartition regime $\epsilon_e = \epsilon_B = 1/3$, are as follows:
the magnetic field $\displaystyle B = 0.052$ G, the~outer radius $\displaystyle R = 3.3\times10^{17}$~cm,
the concentration $n = 1.9$~cm$^{-3}$, the~electron spectral index $s = 3.5$, which is similar to that
obtained from a PIC simulation, and~the minimum energy $\displaystyle E_{1} = 4~m_{e}c^{2}$.

\begin{figure}[H]
	\includegraphics[width=10.5 cm]{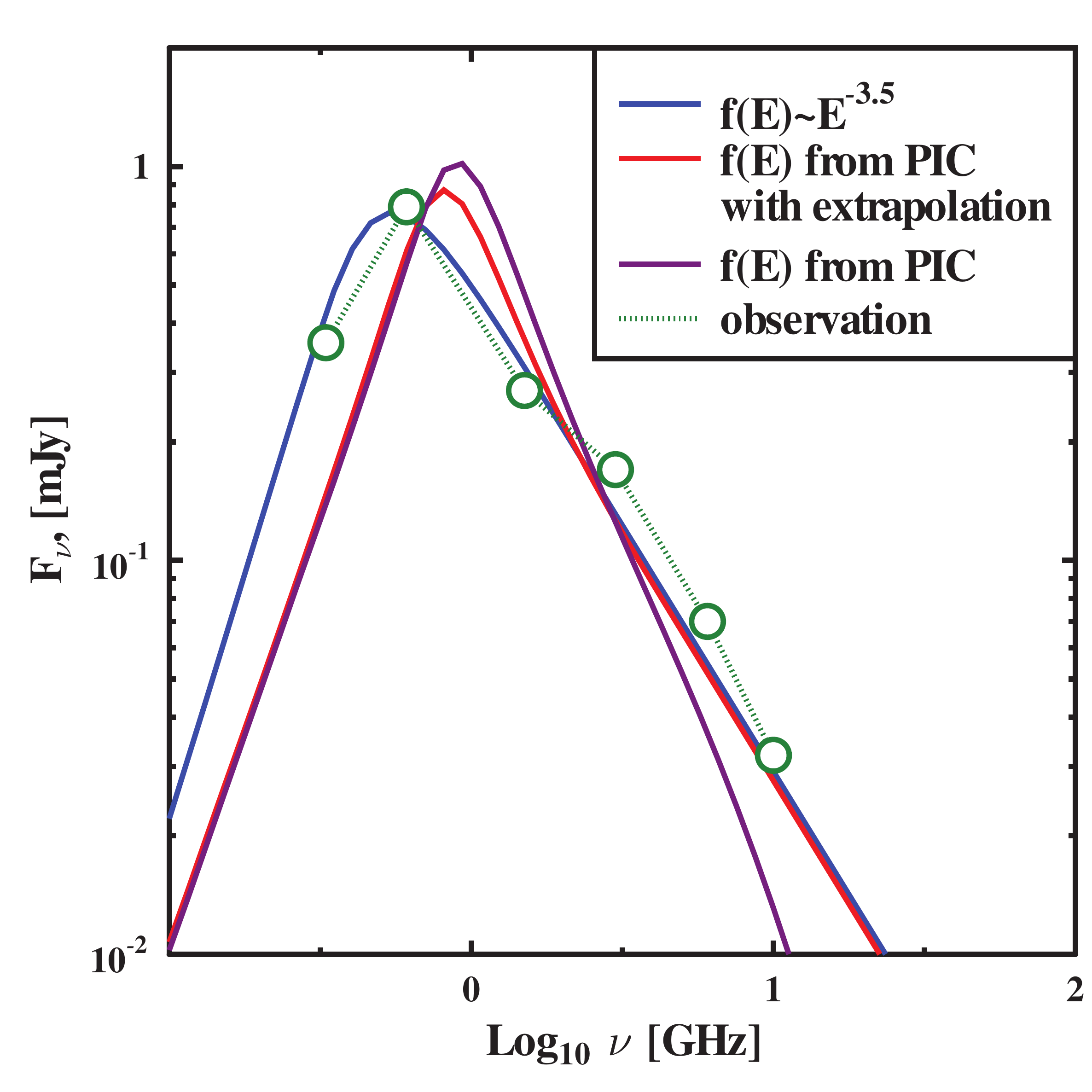}
	\caption{Observed 
 (green circles) and modeled spectral energy distribution of
	CSS161010  at day 357 for various model electron distributions.}
	\label{radiation}
\end{figure}

{The thermal electrons may contribute substantially to the observed  synchrotron spectrum and it was shown in~\cite{Margalit2021} that  models with the power law electron distribution might be oversimplified. Additionally, equipartition regime is not very reasonable assumption. Thus, the~use of the realistic kinetic simulations is important to model broad band radiation spectra of FBOTs. We have compared the observed fluxes to three models of source emission, evaluated with power
law electron distribution and two distributions obtained from PIC simulation: with extrapolation of power law tail to higher energies
($\displaystyle E_{max} = 500~m_{e} c^{2}$) and without it }. For~the first case,
we have used the parameters described above. In~order to evaluate parameters $\displaystyle B$
and $\displaystyle R$ of the simulated particle distribution, we minimized the functional
$\displaystyle g(B,R) = \sum (F(\nu_{i}, B, R) - F_{obs}(\nu_{i}))^{2}$, where $\nu_i$
are the observed frequencies, $\displaystyle F_{obs}(\nu_{i})$ are the observed fluxes and
$\displaystyle F(\nu_{i}, B, R)$---the modeled fluxes, using the gradient descent algorithm.
We employed six measurements from~\cite{2020ApJ...895L..23C}: four made with VLA at day 357
at frequencies 1.5, 3.0, 6.05 and 10~GHz and two made by GMRT at day 350 at frequencies 0.33
and 0.61 GHz. Concentration in these equations is determined by the magnetic field
$\displaystyle\epsilon_{B}$, obtained from the PIC simulation. {One can see that distribution, obtained directly from particle-in-cell simulation, cannot correctly fit the power law tail in observational data, so we had to extrapolate the simulated electron distribution to higher energies
because the PIC approach requires
a lot of computational resources and thus is not suitable to simulate the considered system
up to timescales long enough to form a long tail of accelerated particles.}
The modeled parameters for extrapolated distribution are $\displaystyle B = 0.069$~G, $\displaystyle R = 3.0 \times
10^{17}$~cm and concentration $\displaystyle n = 210$~cm$^{-3}$. The~values of the
magnetic field and radius are rather close to the values from~\cite{2020ApJ...895L..23C}, while the
concentration is much higher. These results
are illustrated in~Figure \ref{radiation}.
\vspace{-12pt}

One can see that the parameters of the shock (especially the concentration),
obtained from the radiation model strongly depend on the electron distribution
function. This should be kept in mind during interpretation of the observational~data.


\section{A Monte Carlo Model of Cosmic Ray Acceleration in Fast Transient~Sources}\label{MC}

We have developed a Monte Carlo model of particle acceleration by collisionless shock waves.
Acceleration occurs according to the first-order Fermi mechanism when particles are scattered
by magnetic fluctuations and cross the shock front many times---the so-called diffusive shock
acceleration (DSA). The~distribution function of the accelerated particles has a significant
anisotropy in the upstream of the shock, and~this anisotropy leads to development of plasma
instabilities in the upstream. The~development of plasma instabilities also leads to a magnetic field
amplification (MFA) in the upstream. The~pressure of the accelerated particles can be on the order
of the total momentum flux flowing onto the shock. In~this case, the~pressure gradient of the
accelerated particles leads to modification of the plasma flow in the~upstream.

The Monte Carlo code employed to describe the DSA and MFA is one-dimensional, stationary, nonlinear
and plane-parallel. We consider acceleration of protons. The~non-relativistic model described in~\cite{BEOV14} has been modified to be applicable for the case of relativistic shocks. Within~this model
all the particles are divided into background and accelerated ones: the accelerated particles are those
that have crossed the shock front from the downstream to the upstream at least once. The~accelerated
particles are treated individually. Background particles are described via macroscopic parameters under
the assumption of their local Maxwell distribution function up to the injection point near the shock
front in the upstream. After~this point, the~background particles are described as particles, which
allows us not to add any additional parameters to describe the injection of particles
into the acceleration process. During~the propagation of particles, they are scattered elastically and
isotropically in the rest frame of the scattering centers according to the pitch-angle scattering
approach~\cite{Jones91,EBJ96,Vladimirov08,Vladimirov09,Vladimirov09dis}. The~reference frame of the
scattering centers for background particles moves with a speed $\displaystyle u\left(x\right)$ relative
to the rest frame of the shock. The~rest frame of the scattering centers for accelerated particles moves
with a speed $\displaystyle u\left(x\right)+v_{scat}\left(x\right)$ relative to the rest frame of the
shock. $\displaystyle u\left(x\right)$ is the speed of the background plasma flow. The~presence of
$\displaystyle v_{scat}\left(x\right)$ is due to the fact that with the development of resonant
instability, the~modes propagating only in a certain direction relative to the background plasma are
amplified. $\displaystyle v_{scat}\left(x\right)$ will be determined~below.

Between the scatterings the particles uniformly move straightforward. The~distance between particle
scatterings is proportional to its mean free path, which we define as:
\begin{equation}\label{lambda}
\lambda\left(x,p\right)=\frac{1}{\frac{1}{\lambda_{B,st}\left(x,p\right)}+\frac{1}{\lambda_{ss}\left(x,p\right)}},
\end{equation}
where $\displaystyle k$ is the absolute value of the wavenumber of modes that make up magnetic
fluctuations, $\displaystyle p$ is the particle momentum, $\displaystyle x$ is the particle coordinate
that is counted from the front of the shock wave. Negative values of $\displaystyle x$ correspond to the
upstream, positive values correspond to the downstream.
\begin{equation}\label{Bohm_st}
\lambda_{B,st}\left(x,p\right)=\frac{pc}{eB_{ls,st}\left(x,k_{res}\right)},
\end{equation}
\begin{equation}\label{B_ls_st}
B_{ls,st}\left(x,k\right)=\sqrt{4\pi\int_{0}^{k}W\left(x,k\right)dk},
\end{equation}
where $\displaystyle W\left(x, k\right)$ is the spectral density of the turbulence energy.
\begin{equation}\label{lambda_ss}
\lambda_{ss}\left(x,p\right)=\left(\frac{pc}{\pi
e}\right)^{2}\frac{1}{\int_{k_{res}}^{\infty}\frac{W\left(x,k\right)}{k}dk},
\end{equation}
\begin{equation}\label{prg_kres}
\frac{k_{res}pc}{eB_{ls}\left(x,k_{res}\right)}=1,
\end{equation}
$B_{ls}$ is the large-scale magnetic field:
\begin{equation}\label{B_ls}
B_{ls}\left(x,k\right)=\sqrt{4\pi\int_{0}^{k}W\left(x,k\right)dk+B_{0}^{2}},
\end{equation}
where $\displaystyle B_{0}$ is the constant longitudinal magnetic field. The~particle propagates until
it gets outside of the model box to the far downstream or crosses the free escape boundary (FEB) located
at $\displaystyle x=x_{FEB}$ in the far~upstream.

The iterative scheme of the employed numerical model allows us to keep the conservation laws of momentum
and energy fluxes near the shock front. In~the stationary relativistic case these are are formulated as
follows. The~law of conservation of particle flux (automatically kept here) has the form:
\begin{equation}\label{Flux_n}
\gamma\left(x\right)\beta\left(x\right)n\left(x\right)=F_{n0},
\end{equation}
where $\displaystyle\gamma\left(x\right)=1/\sqrt{1-\beta^{2}\left(x\right)}$ is the Lorentz factor of
the background plasma flow, $\displaystyle\beta\left(x\right)=u\left(x\right)/c$, $\displaystyle
n\left(x\right)$ is the background plasma number density, $\displaystyle F_{n0}$ is the particle flux in
the far~upstream.

The conservation law of momentum flux takes the form:
\begin{equation}\label{Flux_px}
\gamma^{2}\left(x\right)\beta^{2}\left(x\right)\left[m_{p}c^{2}n\left(x\right)+\Phi_{th}\left(x\right)+\Phi_{w}\left(x\right)\right]+
P_{th}\left(x\right)+P_{w}\left(x\right)+F^{cr}_{px}\left(x\right)=F_{px0}+Q^{ESC}_{px},
\end{equation}
where $P_{th}\left(x\right)$ is the background plasma pressure,
$\displaystyle\Phi_{th}\left(x\right)=\Gamma_{th}\left(x\right)P_{th}\left(x\right)/\left(\Gamma_{th}\left(x\right)-1\right)$, $\Gamma_{th}$ is the background plasma adiabatic index, $P_{w}\left(x\right)$ is
the turbulence pressure, $\displaystyle P_{w}\left(x\right)=0.5\int_{(k)}W\left(x, k\right)dk$,
$\displaystyle\Phi_{w}\left(x\right)=3P_{w}\left(x\right)$, $\displaystyle F_{px0}$ is the momentum flux
in the far unperturbed upstream,  $\displaystyle Q^{ESC}_{px}$ is the momentum flux carried away by the
accelerated particles through the FEB ($\displaystyle Q^{ESC}_{px}=F^{cr}_{px}\left(x_{FEB}\right)$),
$\displaystyle F^{cr}_{px}\left(x\right)$ is the momentum flux of the accelerated~particles.

The conservation law of energy flux takes the form:
\begin{equation}\label{Flux_en}
\gamma^{2}\left(x\right)\beta\left(x\right)\left[m_{p}c^{2}n\left(x\right)+\Phi_{th}\left(x\right)+\Phi_{w}\left(x\right)\right]+F^{cr}_{en}\left(x\right)=F_{en0}+Q^{ESC}_{en},
\end{equation}
where $\displaystyle F_{en0}$ is the energy flux in the far unperturbed upstream, $Q^{ESC}_{en}$ is the
energy flux carried away by accelerated particles through the FEB
($Q^{ESC}_{en}=F^{cr}_{en}\left(x_{FEB}\right)$), $F^{cr}_{en}\left(x\right)$ is the energy flux of the
accelerated~particles.

To keep the conservation laws (\ref{Flux_px}) and (\ref{Flux_en}), it is necessary to determine the
profile of the background plasma flow $\displaystyle u\left(x\right)$ in the upstream and the full
compression by the shock $\displaystyle R_{tot}$ by means of an iterative process. At~the initial
iteration, approximate profiles $\displaystyle u\left(x\right)$, $\displaystyle v_{scat}\left(x\right)$
in the upstream, $\displaystyle W\left(x,k\right)$ and the full shock compression $\displaystyle
R_{tot}$ are set. Then the particles are propagated and their distribution function is calculated
accordingly. In~the far downstream, the~momentum distribution function of all the particles in the rest frame of
the flow is isotropic. Thus, it is possible to determine the adiabatic index of the entire
plasma in the downstream $\displaystyle \Gamma_{p2}$ based on the obtained particle distribution
function. Here and below, the~subscript 0(2) denotes the values in the far upstream (downstream). In~these designations $\displaystyle R_{tot}=\beta_{0}/\beta_{2}$.

To find the full compression, we write down the flux conservation laws (\ref{Flux_n}), (\ref{Flux_px})
and (\ref{Flux_en}) for
the far upstream and for the downstream.
\begin{equation}\label{Flux_n2}
\gamma_{2}\beta_{2}n_{2}=\gamma_{0}\beta_{0}n_{0},
\end{equation}
\begin{multline}\label{Flux_px2}
\gamma_{2}^{2}\beta_{2}^{2}\left[m_{p}c^{2}n_{2}+\Phi_{p2}+\Phi_{w2}\right]+P_{p2}+P_{w2}=\gamma_{0}^{2}\beta_{0}^{2}\left[m_{p}c^{2}n_{0}+\Phi_{th0}+\Phi_{w0}\right]+\\
+P_{th0}+P_{w0}+Q^{ESC}_{px},
\end{multline}
\begin{equation}\label{Flux_en2}
\gamma_{2}^{2}\beta_{2}\left[m_{p}c^{2}n_{2}+\Phi_{p2}+\Phi_{w2}\right]=\gamma_{0}^{2}\beta_{0}\left[m_{p}c^{2}n_{0}+\Phi_{th0}+\Phi_{w0}\right]+Q^{ESC}_{en},
\end{equation}
where $\displaystyle P_{p2}$ is the pressure of the entire plasma in the downstream,
$\displaystyle\Phi_{p2}=\Gamma_{p2}P_{p2}/\left(\Gamma_{p2}-1\right)$. We determine the current values
$\displaystyle Q^{ESC}_{en}$, $\displaystyle Q^{ESC}_{px}$ and $\displaystyle\Gamma_{p2}$ at this
iteration of the quantities from the distribution function obtained after particle propagation. The~value $\displaystyle P_{w2}$ is also calculated based on the equation given below. Thus, three unknowns
$\displaystyle R_{tot}$, $\displaystyle P_{p2}$ and $\displaystyle n_{2}$ remain in the
Equations~(\ref{Flux_n2})--(\ref{Flux_en2}). Solving these equations, we find a new value of $\displaystyle
R_{tot}$. The~value $\displaystyle R_{tot}$ for the next iteration is found by averaging the new value
and the old~one.

A new profile of the flow velocity in the upstream is determined according to the formula based on
(\ref{Flux_px}):
\begin{equation}\label{prof_u}
\gamma_{new}\left(x\right)\beta_{new}\left(x\right)=\gamma_{old}\left(x\right)\beta_{old}\left(x\right)
+\frac{F_{px0}+Q^{ESC}_{px}-F^{cr}_{px}\left(x\right)-F^{th}_{px}\left(x\right)-F^{w}_{px}\left(x\right
)}{\gamma_{0}\beta_{0}m_{p}c^{2}n_{0}},
\end{equation}
where the values obtained after propagation of particles are in the right part of the expression.
The background plasma momentum flux is  $\displaystyle F^{th}_{px}\left(x\right)=\gamma^{2}\left(x\right)\beta^{2}\left(x\right)\Phi_{th}\left(x\right)+P_{th}\left(x\right)$. The~turbulence momentum flux is $\displaystyle
F^{w}_{px}\left(x\right)=\gamma^{2}\left(x\right)\beta^{2}\left(x\right)\Phi_{w}\left(x\right)+P_{w}\left(x\right)$. $\displaystyle \gamma_{old}\left(x\right)\beta_{old}\left(x\right)$ is determined by the flow profile at the previous iteration. Selection of the flow profile based on the expression
(\ref{prof_u}) in the area where the background flow is described in the form of particles, works well
in the case of non-relativistic motion, when calculating in this area $\displaystyle
F^{th}_{px}\left(x\right)$ based on the distribution function of background particles. In~the
relativistic case, as~shown in~\cite{EWB13}, the~momentum flow in the iterative process converges to a
greater value than in the far upstream, when using (\ref{prof_u}) near the shock wave front. Hence,
following~\cite{EWB13} we smooth out the flow profile $\displaystyle u\left(x\right)$ near the shock
front. The~new speed profile is then averaged with the old one. Below~we describe the equations that are
used to calculate the values included in $\displaystyle F^{th}_{px}\left(x\right)$ and
$\displaystyle F^{w}_{px}\left(x\right)$.

The turbulence energy spectrum defining $\displaystyle F^{w}_{px}\left(x\right)$ is
found based on the solution of the following equation:
\begin{multline}\label{eqvFw_k}
\gamma\left(x\right)u\left(x\right)\frac{\partial W\left(x,k\right)}{\partial x}+\frac{3}{2}\frac{\partial\left(\gamma\left(x\right)u\left(x\right)\right)}{\partial x}W\left(x,k\right)
+\frac{\partial \Pi\left(x,k\right)}{\partial k}=\\
=G\left(x,k\right)W\left(x,k\right)-\mathcal{L}\left(x,k\right),
\end{multline}
where $\displaystyle\mathcal{L}\left(x,k\right)$ is the turbulent energy dissipation.
The spectral flux of the turbulent energy (turbulent cascade) is:
\begin{equation}\label{flux_Pi}
\Pi(x,k)=-\frac{C^{*}}{\sqrt{\rho\left(x\right)}}k^{\frac{11}{2}}W\left(x,k\right)^{\frac{1}{2}}\frac{\partial}{\partial k}\left(\frac{W\left(x,k\right)}{k^{2}}\right),
\end{equation}
where $\displaystyle \rho\left(x\right)$ is the background plasma density,
\begin{equation}\label{C_casc}
C^{*}=\frac{3}{11}C_{\rm Kolm}^{-\frac{3}{2}},
\end{equation}
where $\displaystyle C_{\rm Kolm}$ is the Kolmogorov's constant, which is here taken equal to
$\displaystyle C_{\rm Kolm}=1.6$. The~expression (\ref{flux_Pi}) is derived from~\cite{matthaeus2009}.
$\displaystyle G\left(x,k\right)$ is the growth rate of the turbulent energy in the background plasma
rest frame due to plasma instabilities. Similar to~\cite{BEOV14}, here we considered the growth rate of
current instabilities---the Bell's non-resonant~\cite{Bell04} and resonant instability. The~accelerated
particle current, which determines the growth rates, is calculated in the rest frame of the scattering
centers, after~propagation of~particles.

Differentiating the Equations~(\ref{Flux_px}) and (\ref{Flux_en}) by $\displaystyle x$, using
(\ref{Flux_n}), and~excluding from the equations the term proportional to $\displaystyle
n\left(x\right)$ we get the following relation:
\begin{multline} \label{eq_conserv}
\beta\left(x\right)\left(\frac{d\Phi_{th}\left(x\right)}{dx}+\frac{d\Phi_{w}\left(x\right)}{dx}\right)+\frac{dF_{en}^{cr}\left(x\right)}{dx}
+\frac{1}{\gamma\left(x\right)}\frac{d\left(\gamma\left(x\right)\beta\left(x\right)\right)}{dx}\left(\Phi_{th}\left(x\right)+\Phi_{w}\left(x\right)\right)= \\
 =  \beta\left(x\right)\left(\frac{dP_{th}\left(x\right)}{dx}+\frac{dP_{w}\left(x\right)}{dx}+\frac{dF_{px}^{cr}\left(x\right)}{dx}\right).
\end{multline}

If we assume that each of the components of the system is affected only by the change of
$\displaystyle\left(\gamma\left(x\right)\beta\left(x\right)\right)$, that is, the~change in energy and
momentum flux occurs adiabatically due to the change of the flow velocity, we can divide the
Equation~(\ref{eq_conserv}) into separate adiabatic equations for the components:
\begin{equation}\label{eq_conserv_th0}
\beta\left(x\right)\frac{d\Phi_{th}\left(x\right)}{dx}
 +\frac{1}{\gamma\left(x\right)}\frac{d\left(\gamma\left(x\right)\beta\left(x\right)\right)}{dx}\Phi_{th}\left(x\right)=\beta\left(x\right)\frac{dP_{th}\left(x\right)}{dx},
\end{equation}
\begin{equation}\label{eq_conserv_w0}
\beta\left(x\right)\frac{d\Phi_{w}\left(x\right)}{dx}
 +\frac{1}{\gamma\left(x\right)}\frac{d\left(\gamma\left(x\right)\beta\left(x\right)\right)}{dx}\Phi_{w}\left(x\right)= \beta\left(x\right)\frac{dP_{w}\left(x\right)}{dx},
\end{equation}
\begin{equation}\label{eq_conserv_cr0}
\frac{dF_{en}^{cr}\left(x\right)}{dx}= \beta\left(x\right)\frac{dF_{px}^{cr}\left(x\right)}{dx}.
\end{equation}

The Equation~(\ref{eq_conserv_w0}) in this case is equivalent to the Equation~(\ref{eqvFw_k}) integrated
by $\displaystyle k$ in the absence of MFA and dissipation. In~the presence of MFA and dissipation after
integration by $\displaystyle k$, the~Equation~(\ref{eqvFw_k}) will take the form (similar to  (\ref{eq_conserv_w0})):
\begin{multline} \label{eq_conserv_w}
\beta\left(x\right)\frac{d\Phi_{w}\left(x\right)}{dx}
 +\frac{1}{\gamma\left(x\right)}\frac{d\left(\gamma\left(x\right)\beta\left(x\right)\right)}{dx}\Phi_{w}\left(x\right)
 =\beta\left(x\right)\frac{dP_{w}\left(x\right)}{dx}+\\
 +\frac{1}{c\gamma\left(x\right)}\int_{\left(k\right)}G\left(x,k\right)W\left(x,k\right)dk
 -\frac{1}{c\gamma\left(x\right)}L\left(x\right),
\end{multline}
where the dissipation term is  $\displaystyle
L\left(x\right)=\int_{\left(k\right)}\mathcal{L}\left(x,k\right)dk$. Here we use the following
expression for the turbulent energy dissipation:
\begin{equation}\label{eq_diss_x_k}
\mathcal{L}\left(x,k\right)=v_{\Gamma}\left(x\right)\frac{k^{2}}{k_{th}}W\left(x,k\right),
\end{equation}
\begin{equation}\label{v_g}
v_{\Gamma}\left(x\right)=\frac{B_{ls}\left(x,k_{th}\right)}{\sqrt{4\pi\rho\left(x\right)}},
\end{equation}
\begin{equation}\label{prg_k_th}
\frac{k_{th}c\sqrt{k_{b}T_{th}\left(x\right)}}{eB_{ls}\left(x,k_{th}\right)}=1,
\end{equation}
where $\displaystyle k_{b}$ is the Boltzmann's constant, $\displaystyle T_{th}\left(x\right)$ is the
background plasma temperature  ($\displaystyle
P_{th}\left(x\right)=n\left(x\right)k_{b}T_{th}\left(x\right)$).
In this model, it is assumed that in the downstream $\displaystyle\Pi(x,k)=0$ and
$\displaystyle\mathcal{L}\left(x,k\right)=0$.

From the comparison of Equations~(\ref{eq_conserv_w0}) and (\ref{eq_conserv_w}) one can see that
there are two additional terms in the
right side of (\ref{eq_conserv_w}). To~fulfill the total energy conservation
Equation~(\ref{eq_conserv}), these terms must be compensated by introducing additional terms into the equations
for the remaining components of the system. We assume that the dissipation of the turbulent energy flow
leads to heating of the background plasma. An~increase in the energy flow turbulence due to MFA can be
compensated by scattering accelerated particles in the frame of scattering centers moving with the speed
$\displaystyle u\left(x\right)+v_{scat}\left(x\right)$. Then the equation for the energy flux of the
accelerated particles will take the form:
\begin{equation}\label{eq_conserv_cr}
c\frac{dF_{en}^{cr}\left(x\right)}{dx}=[u\left(x\right)+v_{scat}\left(x\right)]\frac{dF_{px}^{cr}\left(x\right)}{dx}.
\end{equation}

Within the considered geometry we introduce $\displaystyle v_{scat}\left(x\right)$ as follows:
\begin{equation}\label{eq_v_scat_full}
v_{scat}\left(x\right)=\min\left(v_{ampl}\left(x\right),v_{A,eff}\left(x\right)\right),
\end{equation}
\begin{equation}\label{eq_v_ampl}
v_{ampl}\left(x\right)=-\frac{\int_{\left(k\right)}G\left(x,k\right)W\left(x,k\right)dk}{\gamma\left(x\right)\frac{dF_{px}^{cr}\left(x\right)}{dx}},
\end{equation}
\begin{equation}\label{eq_v_A_eff}
v_{A,eff}\left(x\right)=-\frac{B_{eff}}{\sqrt{4\pi\rho}},
\end{equation}
\begin{equation}\label{B_eff}
B_{eff}\left(x\right)=\sqrt{4\pi\int_{\left(k\right)}W\left(x,k\right)dk+B_{0}^{2}},
\end{equation}
up to the injection point of the background particles in the upstream. After~the injection point and
before the shock front $\displaystyle v_{scat}\left(x\right)=v_{ampl}\left(x\right)$. In~the downstream
$\displaystyle v_{scat}\left(x\right)=0$.
In this model, we assume that part of the energy flux taken from the accelerated particle flux in the
region where $\displaystyle \left|v_{ampl}\left(x\right)\right|<\left|v_{A,eff}\left(x\right)\right|$,
goes to heat the background plasma. Thus, the~equation determining the change in the energy flux of the
background plasma has the form:
\begin{multline}\label{eq_conserv_th}
\beta\left(x\right)\frac{d\Phi_{th}\left(x\right)}{dx}
 +\frac{1}{\gamma\left(x\right)}\frac{d\left(\gamma\left(x\right)\beta\left(x\right)\right)}{dx}\Phi_{th}\left(x\right)
 = \beta\left(x\right)\frac{dP_{th}\left(x\right)}{dx}+\frac{v_{diss}\left(x\right)}{c}\frac{dF_{px}^{cr}\left(x\right)}{dx}+\\
 +\frac{1}{c\gamma\left(x\right)}L\left(x\right),
\end{multline}
where $\displaystyle v_{diss}\left(x\right)=\left|v_{A,eff}\left(x\right)\right|-\left|v_{ampl}\left(x\right)\right|$ if $\displaystyle \left|v_{ampl}\left(x\right)\right|<\left|v_{A,eff}\left(x\right)\right|$, in~the opposite limit
$\displaystyle v_{diss}\left(x\right)=0$. After~the injection point of background particles and in the
downstream $\displaystyle v_{diss}\left(x\right)=0$.

Substituting the expression for $\displaystyle\Phi_{th}\left(x\right)$ into the
Equation~(\ref{eq_conserv_th}), we obtain the following equation for the background plasma pressure:
\begin{multline}\label{eq_P_th}
\gamma\left(x\right)\beta\left(x\right)\frac{d}{dx}\frac{P_{th}\left(x\right)}{\Gamma_{th}\left(x\right)-1}
+\frac{d\left(\gamma\left(x\right)\beta\left(x\right)\right)}{dx}\frac{\Gamma_{th}\left(x\right)P_{th}\left(x\right)}{\Gamma_{th}\left(x\right)-1}=\\
=\gamma\left(x\right)\frac{v_{diss}\left(x\right)}{c}\frac{dF_{px}^{cr}\left(x\right)}{dx}+\frac{1}{c}L\left(x\right).
\end{multline}

The solution of the Equation~(\ref{eq_P_th}) defines $\displaystyle F^{th}_{px}\left(x\right)$ in the
expression (\ref{prof_u}).

To solve the Equation~(\ref{eqvFw_k}), one needs to set the profile $\displaystyle W\left(x,k\right)$
on the FEB $\displaystyle x=x_{FEB}$. We define $\displaystyle W\left(x_{FEB},k\right)\sim
k^{-\frac{5}{3}}$---the Kolmogorov's spectrum at the FEB with the energy-carrying scale $\displaystyle L_{en}$. $\displaystyle W\left(x_{FEB},k\right)$ is normalized as follows:
\begin{equation}\label{Norm_W}
\int_{\left(k\right)}W\left(x_{FEB},k\right)dk=\frac{B_{0}^{2}}{4\pi}.
\end{equation}

Based on the developed Monte Carlo model, we have performed calculations of proton acceleration by
mildly relativistic shocks, which are thought to come out during explosions of some SN kinds. In~this
case, the~shocks often propagate into the wind of the pre-supernova star. The~number density of
particles (protons) in the stellar wind can be estimated as:
\begin{equation}\label{n_W}
n_{w}\left(r\right)=\frac{\dot{M}}{4\pi v_{w} m_{p} r^{2}},
\end{equation}
where $\displaystyle \dot{M}$ is the mass-loss rate of the pre-supernova,  $\displaystyle v_{w}$ is the
speed of the stellar wind, $\displaystyle r$ is the distance from the center of the star. For~
$\displaystyle \dot{M}=10^{-4} \Msun$yr$^{-1}$, $\displaystyle v_{w}$ = 1000 km s$^{-1}$ and
$\displaystyle r=10^{15}$ cm: $\displaystyle n_{w}\approx 3\times 10^{6}$ cm$^{-3}$. This estimate is
used to estimated the number density $\displaystyle n_{0}$ in the far upstream. $\displaystyle
L_{en}=3\times 10^{16}$~cm in all the calculations. We assume that the free escape boundary is at
0.2 of the current radius of the shock. The~calculation parameters and their results are presented in
Table~\ref{tab_MC}.
\begin{equation}\label{epsilon_B}
\epsilon_{B}'=\frac{B^{2}_{eff,2}}{8\pi\gamma^{2}_{0}u^{2}_{0}m_{p}n_{0}}.
\end{equation}

The spatial coordinate $x$ in the figures is measured in $\displaystyle r_{g0}=m_{p}cu_{0}/eB_{0}$. For~the model A1: $\displaystyle x_{FEB}=-4.782\times 10^{6}r_{g0}$.

\begin{figure}[H]
\includegraphics[width=10.5 cm]{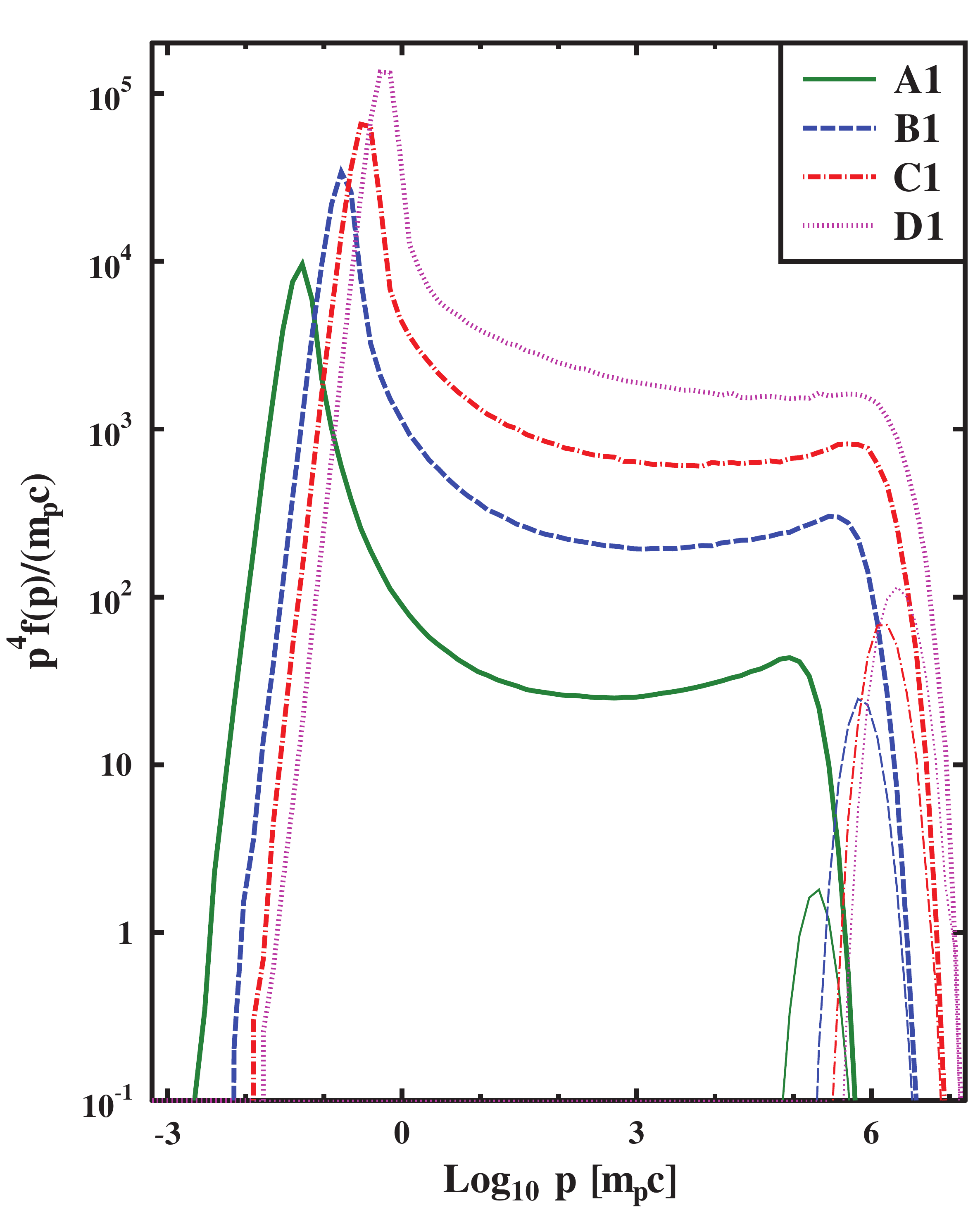}
\caption{Particle 
 distribution function in the shock rest frame, thick curves correspond to the point $\displaystyle x=0$ (the front shock), thin curves correspond to a point $\displaystyle x=x_{FEB}$. The~correspondence of the certain model Table~\ref{tab_MC} is reflected in the~legend.\label{pdf_mid_esc_A1_B1_C1_D1}}
\end{figure}

As can be seen from Figure~\ref{pdf_mid_esc_A1_B1_C1_D1}, with~an increase of the shock speed, the~maximum
momentum of accelerated particles increases. The~value $\displaystyle \epsilon_{B}'$ also increases with
an increase in the velocity of the shock wave {in the range of simulated shock velocities $\displaystyle$ 0.1--0.5 $c$ and it decreases slightly for the shock velocity of $\displaystyle 0.7 c$ (see Table~\ref{tab_MC}) where the transition to relativistic shock regime occurs. Particle-in-cell simulations of electron and proton spectra in the trans-relativistic regime were discussed in~\cite{Crumley, Romansky2018}}.

\begin{table}[H]
\small
\caption{The grid of Monte Carlo calculation~parameters.\label{tab_MC}}
\setlength{\tabcolsep}{3.75mm}
\begin{tabular}{ccccccc}
\toprule
\textbf{Model}	& \boldmath{$u_{0}$}	& \boldmath{$n_{0}$, \textbf{cm}$^{-3}$} & \boldmath{$x_{FEB}$, \textbf{cm}}  & \boldmath{$B_{0}$, \textbf{G}} & \boldmath{$B_{eff,2}$, \textbf{G}} & \boldmath{$\epsilon_{B}'$}\\
\midrule
	A1	& $0.1\times c$ 
	& $5\times 10^{5}$ 
	& $-5\times 10^{14}$ & $3\times 10^{-3}$ & 2.21 & $2.6\times 10^{-2}$\\
	A1n	& $0.1\times c$	& $1\times 10^{5}$	& $-5\times 10^{14}$ & $3\times 10^{-3}$ & 1.00 & $2.7\times 10^{-2}$\\
	A2	& $0.1\times c$	& $2.5\times 10^{3}$	& $-5\times 10^{15}$ & $3\times 10^{-4}$ & $1.57\times 10^{-1}$ & $2.6\times 10^{-2}$\\
	A2b	& $0.1\times c$	& $2.5\times 10^{3}$	& $-5\times 10^{15}$ & $3\times 10^{-3}$ & $1.55\times 10^{-1}$ & $2.5\times 10^{-2}$\\	
	A3	& $0.1\times c$	& $25 $	& $-5\times 10^{16}$ & $3\times 10^{-5}$ & $1.53\times 10^{-2}$ & $2.5\times 10^{-2}$\\	
	B1	& $0.3\times c$	& $5\times 10^{5}$	& $-5\times 10^{14}$ & $3\times 10^{-3}$ & 8.57 & $3.9\times 10^{-2}$\\
	C1	& $0.5\times c$	& $5\times 10^{5}$	& $-5\times 10^{14}$ & $3\times 10^{-3}$ & 16.7 & $4.4\times 10^{-2}$\\
	D1	& $0.7\times c$	& $5\times 10^{5}$	& $-5\times 10^{14}$ & $3\times 10^{-3}$ & 26.3 & $3.8\times 10^{-2}$\\
\bottomrule
\end{tabular}
\end{table}
\vspace{-20pt}

\begin{figure}[H]
\includegraphics[width=10.5 cm]{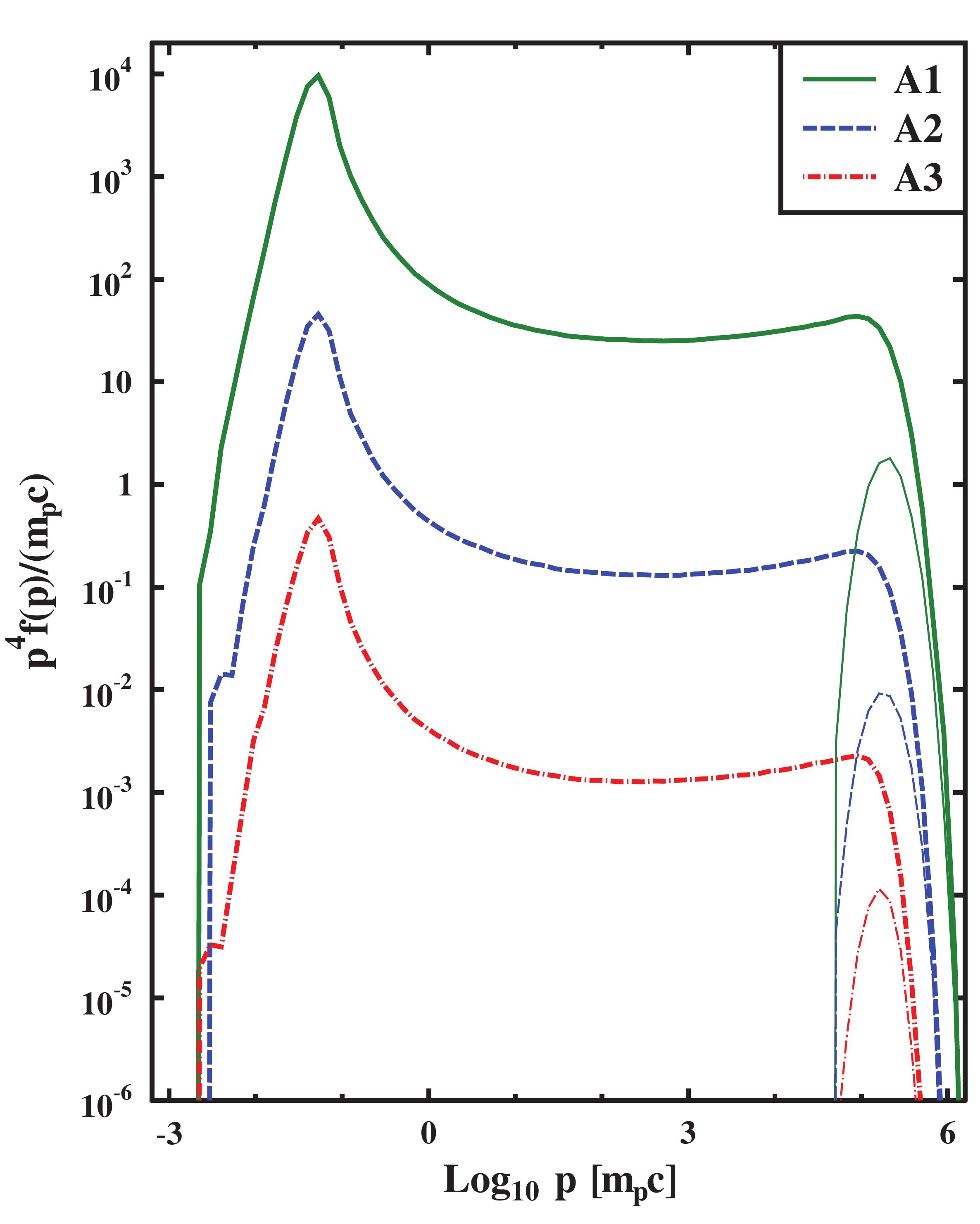}
\caption{The 
 particle distribution function in the shock rest frame, the~thick curves correspond to the
point $\displaystyle x=0$ (the front shock), the~thin curves correspond to the point $\displaystyle
x=x_{FEB}$. The~correspondence to Table~\ref{tab_MC} is reflected in the
legend.\label{pdf_mid_esc_A1_A2_A3}}
\end{figure}

\begin{figure}[H]
\includegraphics[width=10.5 cm]{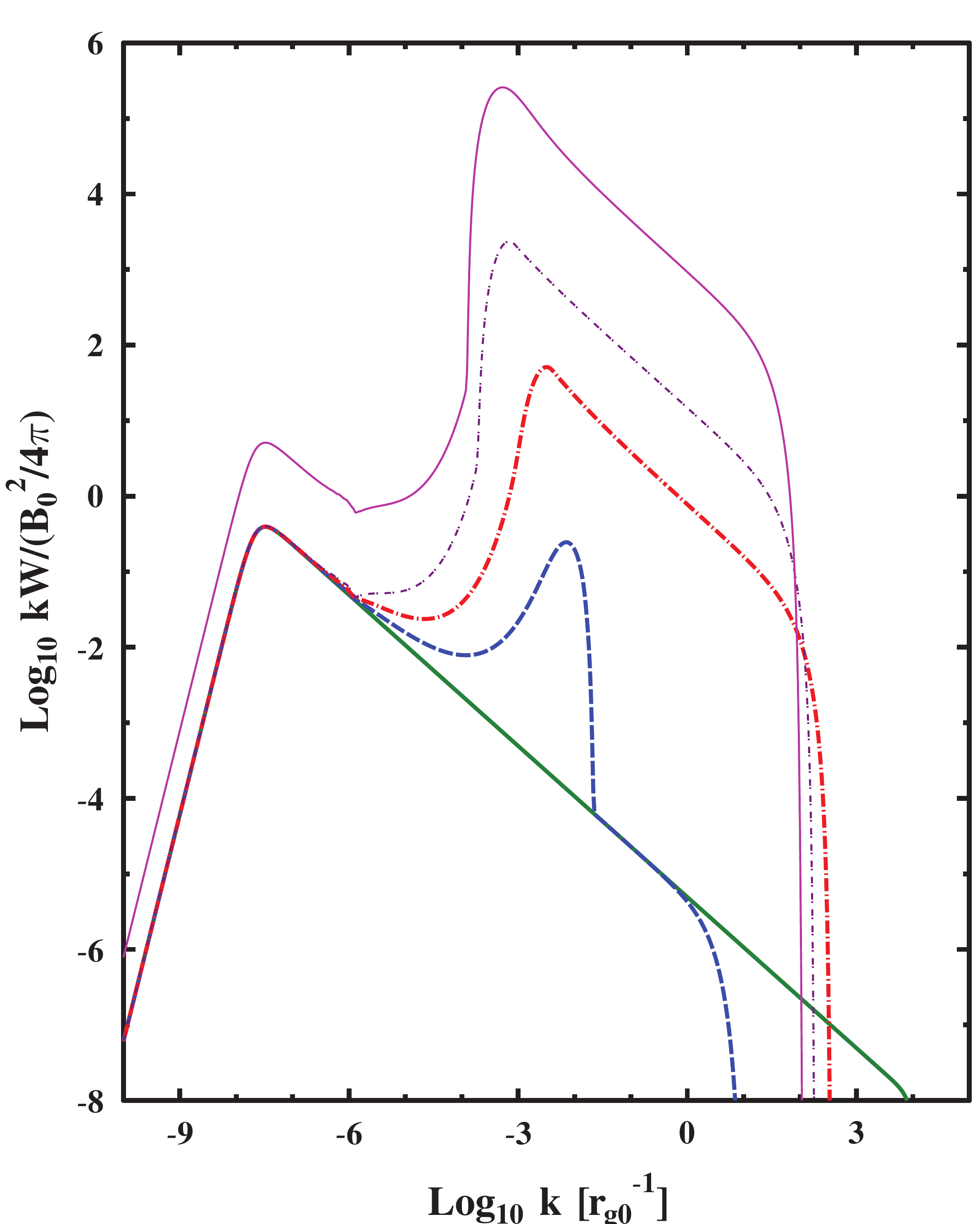}
\caption{Spectral 
 turbulent energy density for the model A1. The~green solid curve corresponds to the point $\displaystyle x=x_{FEB}$. The~blue dashed curve corresponds to the point $\displaystyle x\approx 0.75\cdot x_{FEB}$. The~red dash-point curve corresponds to the point $\displaystyle x\approx 0.5\cdot x_{FEB}$. The~thin purple dashed curve corresponds to the point $\displaystyle x\approx 0.1\cdot x_{FEB}$. The~thin magenta solid curve corresponds to the point $\displaystyle x=0$ (the shock front).\label{W_A1}}
\end{figure}

\begin{figure}[H]
\includegraphics[width=10.5 cm]{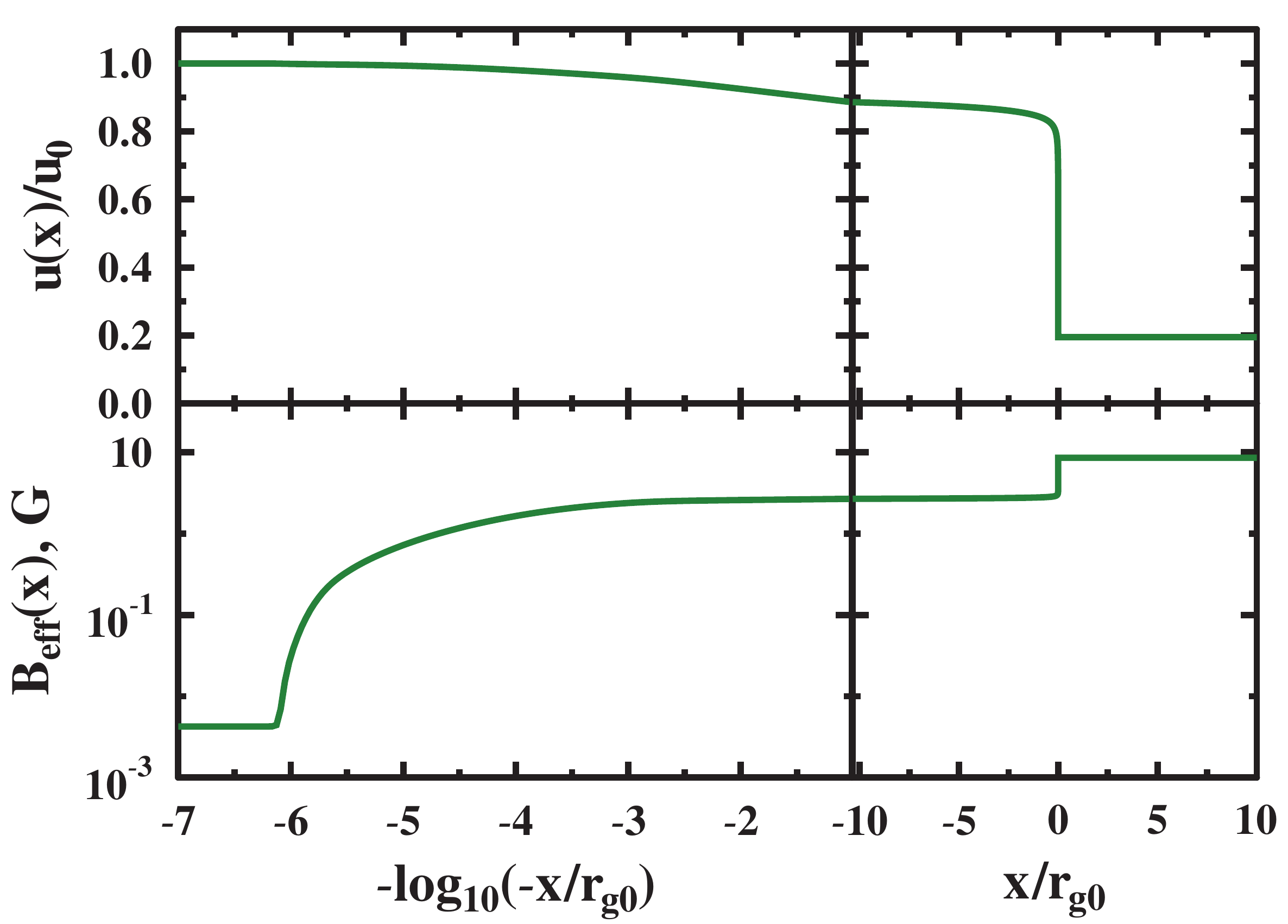}
\caption{Background 
 plasma velocity profile (upper panel) and magnetic field profile (bottom panel) for the model~B1.\label{B_eff_A1}}
\end{figure}

\begin{figure}[H]
\includegraphics[width=10.5 cm]{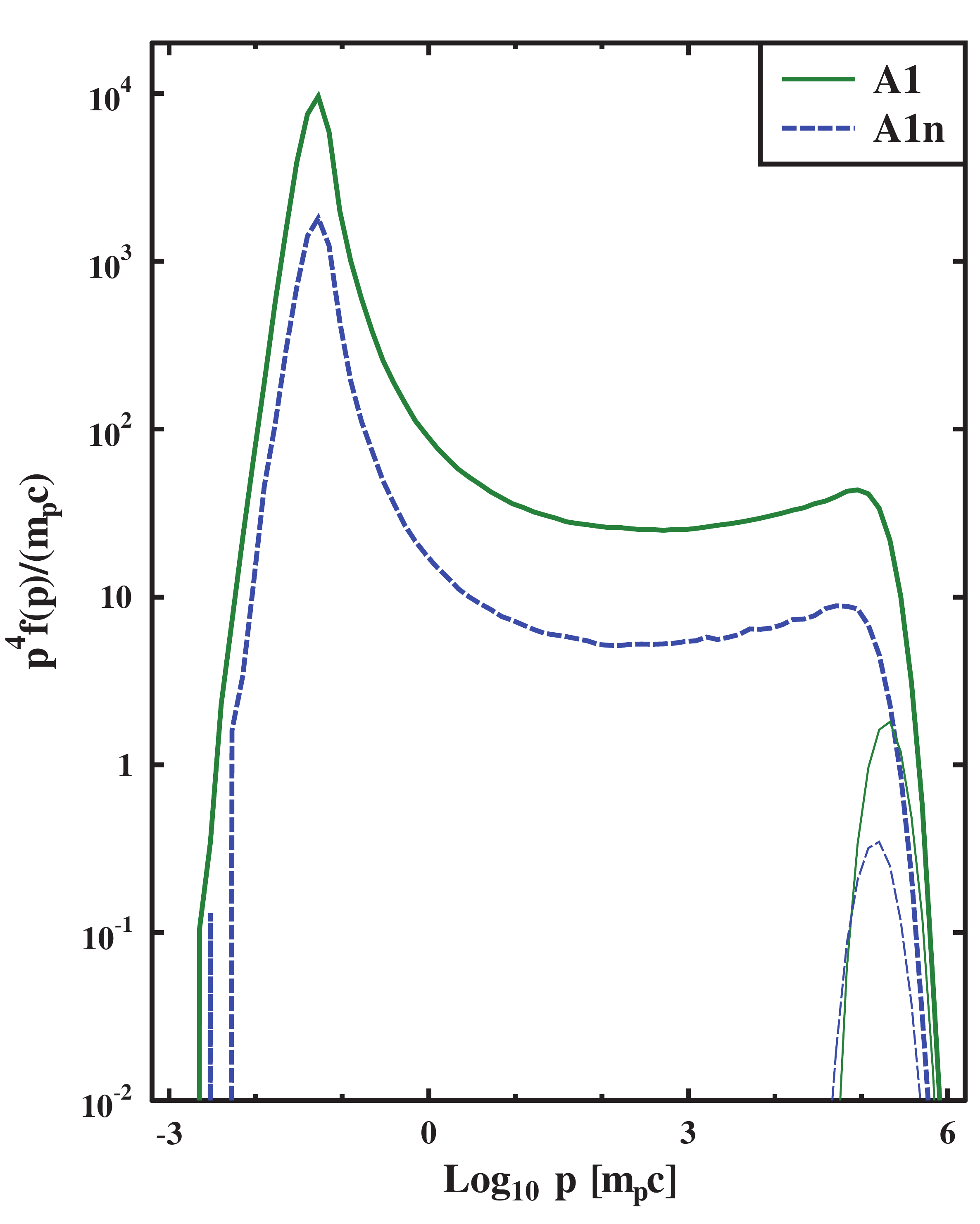}
\caption{Particle 
 distribution function in the shock rest frame, the~thick curves correspond to the
point $\displaystyle x=0$ (the front shock), the~thin curves correspond to the point $\displaystyle
x=x_{FEB}$. The~correspondence to Table~\ref{tab_MC} is reflected in the
legend. \label{pdf_mid_esc_A1_A1n}}
\end{figure}

\begin{figure}[H]
\includegraphics[width=10.5 cm]{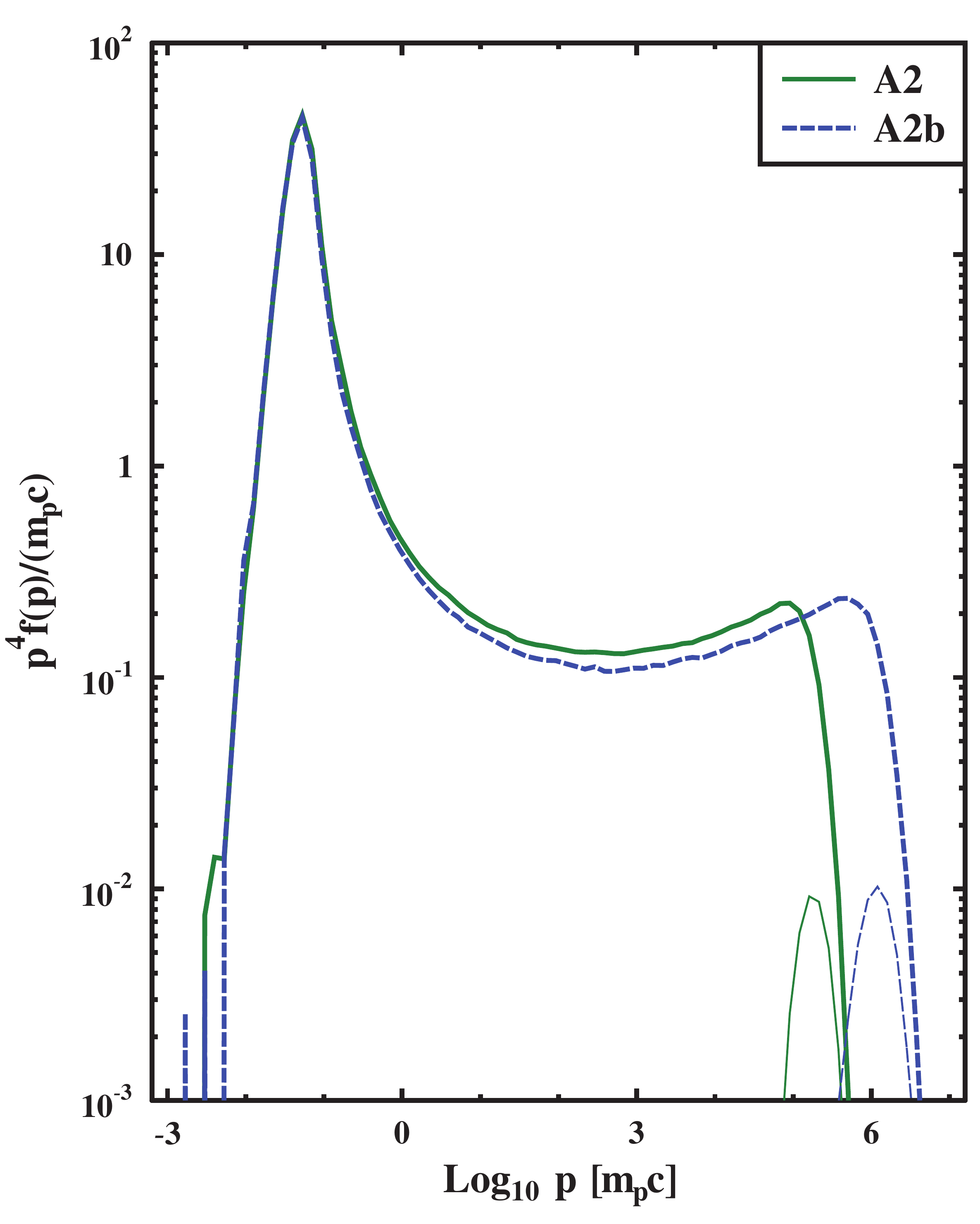}
\caption{Particle 
 distribution function in the shock rest frame, the~thick curves correspond to the
point $\displaystyle x=0$ (the front shock), the~thin curves correspond to the point $\displaystyle
x=x_{FEB}$. The~correspondence to Table~\ref{tab_MC} is reflected in the
legend.\label{pdf_mid_esc_A2_A2b}}
\end{figure}

Figure~\ref{pdf_mid_esc_A1_A2_A3} shows particle distribution functions used for evaluations of the
models A1, A2 and A3, with~varying values of $\displaystyle x_{FEB}$ and the magnetic field at the FEB.
Note, that with an increase of the FEB distance, the~strength of the magnetic field on the FEB decreases
proportionally. It can be seen that in these configurations the maximum momenta of the accelerated
particles are almost the same. The~current of the highest energy particles near the FEB amplify
turbulent fluctuations due to the small-scale Bell's instability on scales much smaller than their own
gyroradius. In~Figure~\ref{W_A1}, the~spectral energy density of the turbulence is shown at various
upstream points for the model A1. Furthermore, it takes time to significantly amplify the magnetic field, and~thus, the~amplitude of the amplified turbulent field in a significant part of the upstream differs
slightly from its value at the FEB (see Figure~\ref{B_eff_A1}). The~gyroradius of the highest energy
particles near the FEB turns out to be significantly larger than the scale of the amplified
fluctuations, which leads to a small contribution of the expression (\ref{lambda_ss}) to the free path
(\ref{lambda}). Thus, the~free path (\ref{lambda}) of the highest energy particles in a
significant part of the upstream is determined by the contribution (\ref{Bohm_st}) with the initial
turbulent field. That is, the~free path (\ref{lambda}) of the highest energy particles in a significant
part of the upstream is $\displaystyle\lambda\left(x,p\right)\sim p/B_{0}$ according to the
normalization (\ref{Norm_W}). The~non-relativistic theory of DSA gives a simple estimate of the maximum
momentum $\displaystyle p_{max}$ in the case of an upstream path independent of the coordinate:
$\displaystyle\lambda\left( p_{max} \right) c/3u_{0}\approx x_{FEB}$. Accordingly, our model has a good
estimate: $\displaystyle p_{max}\sim B_{0}x_{FEB}$. In~Figure~\ref{pdf_mid_esc_A1_A1n}, the~independence
of the maximum particle momentum from the number density  $\displaystyle n_{0}$ in the far upstream is
illustrated. In~Figure~\ref{pdf_mid_esc_A2_A2b}, it is shown, that with an increase of the magnetic field
$\displaystyle B_{0}$ with the other parameters kept constant, the~maximum particle momentum~increases.
\vspace{-13pt}

We have made the following estimate of the acceleration time $\displaystyle\tau_{a}$ of the particles to
their maximum momentum. In~the estimation, we assume that the mean free path of the highest energy
particles with momentum $\displaystyle p_{max}$ in most of the upstream can be estimated by the
gyroradius in the magnetic field $\displaystyle B_{0}$. Thus, taking into account that the magnetic
field in the downstream is much stronger and, accordingly, the~diffusion coefficient of particles is
much smaller than in the upstream $\displaystyle\tau_{a}\approx 3D\left(p_{max}\right)/u_{0}^{2}$, where
$\displaystyle D\left(p_{max}\right)=\lambda\left(p_{max}\right)c/3\approx p_{max}c^{2}/3eB_{0}$ is the
diffusion coefficient of the highest energy particles in the upstream. Accordingly, the~estimate of the
particle acceleration time to the momentum $\displaystyle p_{max}$ has the form:
\begin{equation}\label{tau_accel}
\tau_{a}\approx 3.5\times 10^{5}\left(\frac{u_{0}}{0.1c}\right)^{-2}\left(\frac{B_{0}}{3\times 10^{-3}\textrm{G}}\right)^{-1}\left(\frac{p_{max}}{10^{5}m_{p}c}\right)\textrm{s}.
\end{equation}

\vspace{-20pt}

In the considered models we assume that the current radius of the shock is $\displaystyle
R_{f}=5\left|x_{FEB}\right|$. The~expansion time of the supernova remnant can be estimated as the ratio
of the current shock radius to the shock velocity. Accordingly, the~estimate for the expansion time has
the form:
\begin{equation}\label{tau_exp}
\tau_{exp}\approx 8.3\cdot 10^{5}\left(\frac{\left|x_{FEB}\right|}{5\cdot 10^{14}\textrm{cm}}\right)\left(\frac{u_{0}}{0.1c}\right)^{-1}\textrm{s}.
\end{equation}

It can be seen from the expressions (\ref{tau_accel}), (\ref{tau_exp})  that the expansion time is
longer than the acceleration time for the model (see Table~\ref{tab_MC}), the~maximum momentum can
be estimated from Figure~\ref{pdf_mid_esc_A1_B1_C1_D1}, which confirms the self-consistency of the
stationary model we have~employed.

\vspace{-20pt}

\unskip

\section{Discussion and~Conclusions}

In order to understand the processes of non-thermal particle acceleration at work in fast and
energetic transients we have carried out and presented above a microscopic gyro-scale modeling of the
collisionless shock structure and non-thermal particle spectra, which show a strong dependence of both
electron and proton acceleration efficiency on the shock obliquity.

{In this context one should keep in mind that the particle-in-cell simulation assumed a fixed homogeneous
magnetic field and cold flow of the incoming particles in the shock upstream at the boundary of the simulation box. On~the other hand, fast non-thermal particles can penetrate into the upstream plasma flow. It is clear from the results of the Monte Carlo modeling presented in Section \ref{MC} (see, in~particular, Figure~\ref{B_eff_A1}) that the
efficient cosmic ray acceleration at the shock may provide a strong magnetic field amplification and modification of the upstream plasma flow. The~cosmic ray pressure gradient in the shock upstream decelerates the incoming plasma flow and the cosmic ray driven instabilities may highly amplify the fluctuating magnetic fields well outside the
particle-in-cell simulation box which is limited by a few hundred of the proton gyroradii around the plasma sub-shock. Such simulations show the potential importance of the feedback effects
which cannot be modeled with the microscopic particle-in-cell simulations so far.  Therefore we have used here the particle-in-cell model to simulate the electron injection and acceleration in sub and mildly relativistic shocks
of fast energetic transients together with Monte Carlo modeling. While the Monte Carlo technique describes the structure of upstream flow modified by the accelerated particles at scales well above the proton gyro-scales, it cannot be used to simulate the electron injection where particle-in-cell approach is required.    Thus, we used the combination of the two techniques to model radio emission.}

{In our study, of~electron spectrum convergence in quasi-parallel shock of $0.3 c$ speed we observed some non-monotonic temporal behavior of the electron spectra as it is shown in Figure~\ref{electrons_time}. This could be due to the development of Bell’s instability as it was found earlier in~\cite{Crumley} and which mediate the quasi-parallel shock structure at proton gyro-scales. The~maximum electron energy achievable in our simulation was about 50 $m_e c^2$ and we extrapolate the spectral slope to larger energy. We extended the spectra from 50 to 500 $m_e c^2$ using the same spectral slope, it is enough to model radio spectra. This is an assumption in our case. However, the~extrapolation seems to be justified by the presence of a power-law such as a spectral component with a similar slope  right after the thermal peak  in Crumley~et~al. simulation (see the electron spectra presented in Figure~7 
~\cite{Crumley} in the electron momentum range between 0.1 and 30 $m_p c$ ). A~similar spectral component is also apparent right after the peak of the spectrum in our steady Monte Carlo simulations (see, e.g.,~our  Figure~\ref{pdf_mid_esc_A1_A2_A3}).

We applied 2D kinetic PIC simulations for sub-relativistic flows with velocity of 0.3~c 
 to model the spectra of electrons and protons. The~dynamical range of the full kinetic PIC simulations is limited but the electron spectrum extrapolated with account for the results of Monte Carlo simulations is enough to model the observed radio emission in the fast transients where such fast outflows were found. In~the Monte Carlo model the particle-in-cell simulated domain where the electron injection occurs correspond to the sub-shock structure which is apparent at $\displaystyle x=0$ 
 in Figure~\ref{B_eff_A1}. The~plasma compression ratio at the sub-shock of the  cosmic ray modified shock is about 3 and the spectra of particles accelerated at the sub-shock in the low energy regime (c.f. the proton spectral slope right after the peak in    Figure~\ref{pdf_mid_esc_A1_A2_A3}) are consistent with the ones obtained within the particle-in-cell simulation dynamical range. The~gyroradius of electrons with Lorentz-factor  of $\displaystyle 1000$ in our simulations is about the gyroradius of protons with Lorentz-factor of  $\displaystyle 10$. The~spectral index of proton distribution simulated with Monte Carlo model at Lorentz-factor of $\displaystyle 10$ is consistent with the PIC electron spectrum extrapolated to $\sim$500 $m_e c^2$.}

This allows to model the non-thermal emission and to understand the
origin of the synchrotron radio emission of FBOTs. Additionally, one may assume an explanation of their hard X-ray spectrum using the dependence of the electron power law distribution index on the shock obliquity. As~can be seen in Figure~\ref{compareelectrons} it leads
to the presence of relativistic electron populations of comparable intensities within a range of spectral
indices along the curved shock surface expected in a wide angle outflow. In~this case the synchrotron
radio emission could have rather steep spectral indices, while the X-ray component produced by
the inverse Compton scattering of the intense optical radiation by the relativistic electrons with the
harder spectral index may appear to be~flatter.

The simplified macroscopic Monte Carlo kinetic model was used in Section \ref{MC} to estimate the maximum
energies of the accelerated cosmic ray nuclei. The~model accounted for both the nonlinear feedback
effects of the flow modification by the cosmic ray pressure gradient and magnetic turbulence
amplification by cosmic ray driven instabilities in the upstream of a plain mildly relativistic shock.
The model demonstrated a possibility of PeV regime proton acceleration in the shocks driven by
mildly relativistic outflows of fast energetic transients on a few weeks~timescale.

Transient objects of different types such as gamma-ray bursts and  tidal disruption events~\cite{2014arXiv1411.0704F,2016PhRvD..94j3006M,2020ApJ...902..108M}  are widely discussed as possible
sources of the observed high energy neutrinos. The~search of PeV gamma-ray sources have recently
revealed a number of potential galactic pevatron sources (see, e.g.,
\cite{2019NatAs...3..561A,2021PhRvL.126n1101A,LHAASONat21,2021ApJ...916L..22D}). The~analysis presented
in~\cite{2021Univ....7..324C} allowed the authors to conclude that the extended galactic supernova
remnants are not likely to be pevatrons, while some other types of galactic
sources associated with the SN events in the compact clusters of young massive stars
~\cite{2015MNRAS.453..113B} and gamma-ray binaries with compact relativistic stars
~\cite{2021ApJ...915...31K,2021arXiv211011189B} can produce PeV photons and neutrinos.
While PeV photons undergo strong attenuation and can be detected mostly from the galactic
sources, high energy neutrinos from extragalactic energetic transients can be detected with currently
operating and future neutrino observatories simultaneously with LSST optical transients~detection.





\acknowledgments{We are honored to dedicate this paper to the memory of professor Yury~ Nikolaevich~Gnedin
(1935--2018). His great scientific enthusiasm, competence and the unfailing goodwill for many years were
extremely important for us. We are grateful to the referees for constructive comments and suggestions.  The~authors  were supported by the RSF grant 21-72-20020. Some of the
modeling was performed at the Joint Supercomputer Center JSCC RAS and some---at the ``Tornado''
subsystem of the Peter the Great Saint-Petersburg Polytechnic University Supercomputing Center.}

\reftitle{References}



\end{paracol}
\end{document}